\documentclass[preprint,prd,nofootinbib,tightenlines,amsmath,amssymb]{revtex4}
\usepackage{enumerate}
\usepackage{multirow}
\usepackage[utf8]{inputenc}
\usepackage{anysize}
\usepackage{textcomp}
\usepackage{epsfig}
\usepackage{graphics,color} 
\usepackage{verbatim}
\usepackage{afterpage}
\usepackage{amsmath}    
\usepackage{soul}
\usepackage{xcolor,cancel}
\DeclareUnicodeCharacter{00A0}{ }

\usepackage{blindtext}

\catcode`\@=11
\def\sla@#1#2#3#4#5{{%
 \setbox\z@\hbox{$\m@th#4#5$}%
 \setbox\tw@\hbox{$\m@th#4#1$}%
 \dimen4\wd\ifdim\wd\z@<\wd\tw@\tw@\else\z@\fi
 \dimen@\ht\tw@
 \advance\dimen@-\dp\tw@ \advance\dimen@-\ht\z@
 \advance\dimen@\dp\z@
 \divide\dimen@\tw@ \advance\dimen@-#3\ht\tw@
 \advance\dimen@-#3\dp\tw@ \dimen@ii#2\wd\z@
 \raise-\dimen@\hbox to\dimen4{%
 \hss\kern\dimen@ii\box\tw@\kern-\dimen@ii\hss}%
 \llap{\hbox to\dimen4{\hss\box\z@\hss}}}}

\def\cpto{\mathrel {\vcenter {\baselineskip 0pt \kern 0pt
    \hbox{$H_{r.f.}$} \kern 0pt \hbox{$\longrightarrow$} }}}
\def\slashed#1{%
 \expandafter\ifx\csname sla@\string#1\endcsname\relax
{\mathpalette{\sla@/00}{#1}}
\fi}
\def\declareslashed#1#2#3#4#5{%
 \expandafter\def\csname sla@\string#5\endcsname{%
#1{\mathpalette{\sla@{#2}{#3}{#4}}{#5}}}}
 \catcode`\@=12
\declareslashed{}{/}{.08}{0}{D}
 \declareslashed{}{/}{.1}{0}{A}
 \declareslashed{}{/}{0}{-.05}{k}
 \declareslashed{}{/}{.1}{0}{\partial}
 \declareslashed{}{\not}{-.6}{0}{f}

\def\lsim{\mathrel {\vcenter {\baselineskip 0pt \kern 0pt
    \hbox{$<$} \kern 0pt \hbox{$\sim$} }}}
\def\gsim{\mathrel {\vcenter {\baselineskip 0pt \kern 0pt
    \hbox{$>$} \kern 0pt \hbox{$\sim$} }}}

\newcommand{\bea}{\begin{eqnarray}}
\newcommand{\eea}{\end{eqnarray}}

\begin{document}

\baselineskip=15pt
\preprint{}

\title{LHC constraints on color octet scalars}

\author{Alper Hayreter$^1$\footnote{Electronic address: alper.hayreter@ozyegin.edu.tr},  German Valencia$^{2}$\footnote{Electronic address: German.Valencia@monash.edu }}
\affiliation{
$^{1}$Department of Natural and Mathematical Sciences, Ozyegin University, 34794 Istanbul Turkey.\\
$^{2}$School of Physics and Astronomy, Monash University, Wellington Road, Melbourne, Victoria 3800, Australia}

\date{\today}

\vskip 1cm
\begin{abstract}

We  extract constraints on the parameter space of the MW model by comparing the cross-sections for dijet, top-pair, dijet-pair, $t\bar t t \bar t$ and $b\bar b b \bar b$ productions at the LHC with the strongest available experimental limits from ATLAS or CMS at 8 or 13 TeV. Overall we find mass limits around 1~TeV in the most sensitive regions of parameter space, and lower elsewhere. This is at odds with generic limits for color octet scalars  often quoted in the literature where much larger production cross-sections are assumed. The constraints that can be placed on coupling constants are typically weaker than those from existing theoretical considerations, with the exception of the parameter $\eta_D$.

\end{abstract}

\pacs{PACS numbers: }

\maketitle

\section{Introduction}

The search for new physics at the LHC has covered enormous ground so far, constraining the parameters of many possible new particles that appear in a variety of extensions of the standard model (SM). There are still exceptions, models that have not been carefully confronted with the LHC data, amongst them an extension of the scalar sector of the SM with  a color octet electroweak doublet scalar. This model was introduced some time ago by Manohar and Wise (MW) \cite{Manohar:2006ga}, and it is motivated by minimal flavor violation \cite{Manohar:2006ga}.

There are several phenomenological papers concerning LHC studies of the MW model available, but all of them predate the LHC. The original MW paper already gave analytic expressions for the pair production of the new scalars through the dominant gluon fusion channel, finding production cross-sections that vary many orders of magnitude, between $~50$~fb and $10^{-3}$~fb for $M_S$ between $1$ and $ 3$~TeV respectively at LHC14 \cite{Manohar:2006ga}. Shortly after, Gresham and Wise considered the production of a single neutral scalar, a one loop process similar to the production of the Higgs boson, which becomes dominant over the two scalar production at around 1~TeV \cite{Gresham:2007ri}. Gerbush {\it et al} \cite{Gerbush:2007fe} considered pair production of scalars leading to final states with four heavy quarks (top or bottom) at the LHC concluding that discovery would be possible for masses up to about a TeV. Finally Arnold and Fornal considered constraints that can be imposed by studying high $p_T$ four-jet events at LHC \cite{Arnold:2011ra}, and concluded that 10~fb$^{-1}$ would be enough to discover a scalar as heavy as 1.5~TeV at LHC14.

In this paper we partially address the existing gap in the extraction of LHC constraints for the MW model by comparing the cross section for production of a single neutral scalar and a pair of neutral scalars to limits obtained by ATLAS and CMS in the dijet, top-pair, dijet-pair, $t\bar t t \bar t$ and $b\bar b b \bar b$ channels. We defer the study of charged scalar production and their different decay modes to a future publication. We restrict this paper to a parton-level study because its main weakness is that there is no consistent set of LHC data that can be applied to this model.

A similar study for pseudo-scalar color octets, common to composite Higgs models, exists in the literature \cite{Belyaev:2016ftv}.

\section{The Model}

The MW model contains many new parameters, several of which have been studied phenomenologically before. In particular the new color octet scalars can have a large effect on loop level Higgs production and decay \cite{Manohar:2006ga} and this has resulted in  multiple studies of  one-loop effective Higgs couplings \cite{He:2011ti,Dobrescu:2011aa,Bai:2011aa,Cacciapaglia:2012wb,Cao:2013wqa,He:2013tia}. They are also constrained by precision electroweak measurements \cite{Manohar:2006ga,Gresham:2007ri,Burgess:2009wm},  flavor physics \cite{Grinstein:2011dz,Cheng:2015lsa,Martinez:2016fyd}, unitarity and vacuum stability \cite{Reece:2012gi,He:2013tla,Cheng:2016tlc,Cheng:2017tbn} and other LHC processes \cite{Gerbush:2007fe,Arnold:2011ra,He:2011ws,Kribs:2012kz}.

In the MW model, the new field $S$ transforms as $(8,2,1/2)$ under the SM gauge group $SU(3)_C\times SU(2)_L\times U(1)_Y$ and this gives rise to the gauge interactions responsible for its pair production at the LHC. Numerous new couplings appear in the self-interactions of the scalars as well as in the Yukawa couplings to fermions. The latter are restricted by minimal flavor violation to two complex numbers \cite{Manohar:2006ga}, 
\begin{equation}
{\cal L}_Y=-\eta_U e^{i\alpha_{U}} g^{U}_{ij}\bar{u}_{Ri}T^A Q_j S^A - \eta_D e^{i\alpha_{D}}g^{D}_{ij}\bar{d}_{Ri}T^A Q_j S^{\dagger A} + h.c,
\label{yukc}
\end{equation}
where $Q_i$ are left-handed quark doublets, $S=S^aT^a \;(a=1,...,8)$ and the $SU(3)$ generators are normalized as ${\rm Tr} (T^aT^b)=\delta^{ab}/2$. The  matrices $g^{U,D}_{ij}$ are the same as the  Higgs couplings to quarks, and $\eta_{U,D}$ along with their phases $\alpha_{U,D}$, are new overall factors. Non-zero phases signal $CP$ violation beyond the SM and contribute for example to the EDM and CEDM of quarks \cite{Manohar:2006ga,Burgess:2009wm,He:2011ws,Martinez:2016fyd}. 

The most general renormalizable scalar potential is given in Ref.~\cite{Manohar:2006ga} and contains several terms. Of these, our study will only depend on the following (with $v\sim 246$~GeV the usual Higgs vacuum expectation value)
\begin{eqnarray}
&&V=\lambda\left(H^{\dagger i}H_i-\frac{v^2}{2}\right)^2+2m_s^2\ {\rm Tr}S^{\dagger i}S_i +\lambda_1\ H^{\dagger i}H_i\  {\rm Tr}S^{\dagger j}S_j +\lambda_2\ H^{\dagger i}H_j\  {\rm Tr}S^{\dagger j}S_i 
\nonumber \\
&&+\left( \lambda_3\ H^{\dagger i}H^{\dagger j}\  {\rm Tr}S_ iS_j +\lambda_4\ e^{i\phi_4}\ H^{\dagger i} {\rm Tr}S^{\dagger j}S_ jS_i +
\lambda_5\ e^{i\phi_5}\ H^{\dagger i} {\rm Tr}S^{\dagger j}S_ iS_j 
+{\rm ~h.c.}\right)
\label{potential}
\end{eqnarray}
The number of parameters in Eq.(\ref{potential}) can be reduced for our study as follows: first $\lambda_3$ can be chosen to be real by a suitable definition of $S$; then custodial $SU(2)$ symmetry can be invoked to introduce the relations $2\lambda_3=\lambda_2$ (and hence $M_{S^+}=M_{S_I}$) \cite{Manohar:2006ga} and $\lambda_4=\lambda_5^\star$ \cite{Burgess:2009wm}; and finally, requiring  CP conservation removes all the phases, $\alpha_U$, $\alpha_D$, and  $\phi_4$.

After symmetry breaking, the non-zero vev of the Higgs field in Eq.(\ref{potential}) gives the physical $h$ its usual mass $m^2_H = 2 \lambda v^2$, and in addition it  splits  the octet scalar masses as,
\begin{eqnarray}
m^2_{S^{\pm}} =  m^2_S + \lambda_1 \frac{v^2}{4},&&
m^2_{S^{0}_{R,I}} =  m^2_S + \left(\lambda_1 + \lambda_2 \pm 2 \lambda_3 \right) \frac{v^2}{4},
\end{eqnarray}
These relations, combined with the use of custodial and CP symmetries result in the following independent input parameters:
\begin{eqnarray}
M_{S_R},\ \lambda_2,\ \lambda_4,\ \eta_U,\ \eta_D
\end{eqnarray}
The parameter $\lambda_2$ controls the split between the two neutral resonances $S_{I,R}$ and the parameter $\lambda_4$ controls the strength of scalar loop contributions to single neutral scalar production through gluon fusion. The parameters $\eta_{U,D}$ control respectively the strength of the $Stt$ and $Sbb$ interactions. 

Under the above assumptions, the effective one-loop  coupling $Sgg$ can be written in terms of two factors $F_{R,I}$ as,
\begin{eqnarray}
{\cal L}_{Sg g }&=& F_R G^A_{\mu\nu} G^{B\mu\nu} d^{ABC}S^{C0}_R+ F_I \tilde G^A_{\mu\nu}G^{B\mu\nu}d^{ABC}S^{C0}_I\;,
\end{eqnarray}
where $G^A_{\mu\nu}$ is the gluon field strength tensor and $\tilde G^{A\mu\nu} = (1/2)\epsilon^{\mu\nu\alpha\beta}G^A_{\alpha\beta}$.

At one-loop level these factors receive their main contributions from top-quark and scalar loops and are given by \cite{Gresham:2007ri}
\begin{eqnarray}
F_R &=& (\sqrt{2}G_F)^{1/2} \frac{\alpha_s}{8\pi}\left[ \eta_U \, I_q\left(\frac{m^2_t }{m_R^2}\right)+\eta_D \, I_q\left(\frac{m^2_b }{m_R^2}\right) \right. \nonumber \\
&-& \left. \frac{9}{ 4} \frac{v^2}{m^2_R} \lambda_4  \left(I_s(1)+I_s\left(\frac{m_I^2}{m_R^2}\right)\right)\right]\;,\nonumber\\
F_I &=& (\sqrt{2}G_F)^{1/2} \frac{\alpha_s}{8\pi}\frac{1}{2} \left[\frac{m^2_t
 }{ m^2_I} \eta_U \, f\left(\frac{m^2_t}{m_I^2}\right)+\frac{m^2_b
}{ m^2_I} \eta_D \, f\left(\frac{m^2_b}{m_I^2}\right)\right].
\label{loopc}
\end{eqnarray}
We have already simplified Eq.(\ref{loopc}) with the relations $\lambda_5=\lambda_4$, $m^\pm = m_I$, and we have allowed for the mass  $m_R$, to be different from $m_I$ through $\lambda_2$ as discussed above. Throughout the calculation, the scalars $S_R$ and $S_I$ (when not in a loop), are taken to be on-shell, in keeping with the narrow width approximation.
The loop functions $I_{q,s}$  and $f$ are well known and given by:
\begin{eqnarray}
I_q(z) &=& 2z+z(4z-1)f(z), \quad 
I_s(z) = -z(1+2zf(z)), \nonumber \\
f(z) &=& \frac{1}{2}\left(\ln\left(\frac{1+\sqrt{1-4z}}{1-\sqrt{1-4z}}\right)-i\pi\right)^2 \ \ {\rm for~}z<1/4 \nonumber \\
&=&  -2 \left(\arcsin\left(\frac{1}{2\sqrt{z}}\right)\right)^2 \ \ {\rm for~}z>1/4 .
\label{loopintegrals}
\end{eqnarray}
From this it follows that the top-quark contribution starts as large as $I_q(1/4) = 1/2$ for a resonance with  twice the top-quark mass, and goes to zero as $m_S >> m_t$. Additionally, the factor 
$I_s(1)=\pi^2/9-1$,  generates some suppression in contributions from scalar loops relative to top-quark loops. For this reason only relatively large values of $\lambda_4$ make a significant contribution to single scalar production. We have also included the bottom-quark loop, which is important only for regions of parameter space where $|\eta_D|>>|\eta_U|$. This situation is completely analogous to Higgs production in type II two Higgs doublet models where the $b$ quark contribution is important for $\tan\beta>>1$. We will consider this case  for the dijet final state study.

\section{Confronting LHC searches}

An early paper constraining new particles at the LHC introduced several benchmarks \cite{Han:2010rf}, including a color octet scalar $S$, that have  since been pursued by both ATLAS \cite{Aad:2014aqa} and CMS  \cite{Sirunyan:2016iap}.  Both these studies appear to rule out the new color octet scalars with masses up to 3~TeV. Unfortunately, the original benchmark  does not cover interesting possibilities for additional color octet scalars like the MW model. This can be easily seen from their coupling to two gluons,  
$F_R=g_s \frac{\kappa_S}{\Lambda_S}$, 
which has been used with $\kappa_S=1/\sqrt{2}$ in the CMS study. Comparing this to $F_R$ in Eq.(\ref{loopc}), one sees that realistic couplings in the MW model result in cross-sections that are three orders of magnitude below the  benchmark.

To proceed with our numerical study we implement the Lagrangian of  Eq.(\ref{yukc})~and~Eq.(\ref{loopc}) in FEYNRULES \cite{Christensen:2008py,Degrande:2011ua} to generate a Universal Feynrules Output (UFO) file, and then feed this UFO file into MG5\_aMC@NLO \cite{Alwall:2014hca}.  To extract constraints on the parameters of the new resonances we will compare our predicted cross-sections times branching ratios with existing LHC studies. We divide our study according to the different channels that dominate different regions of parameter space. We will use in particular the following channels: two jets; top pairs; four jets; four top-quarks; and four bottom quarks. 

We illustrate all cases with production and decay of $S_R$ or $S_R S_R$ but the limits are identical for  $S_I S_I$ and similar for $S_I$ provided we  prevent the decays $S_{R,I}\to S^\pm W^\mp j$ (which is done by a choice of $\lambda_2$). If those decays are allowed, different limits apply from channels involving final state leptons which we do not consider in this paper. In the limit of $CP$ conservation that we consider here, the $S_I$ is a pseudo-scalar and therefore its gluon fusion production mechanism is a bit different from $S_R$, being independent of $\lambda_4$.

\subsection{Search for resonances in dijets at LHC}

This channel has been studied before in the context of the MW model for the Tevatron. A comparison with the dijet cross-section at the Tevatron revealed that signatures from the MW model are orders of magnitude below the QCD background and no constraint emerged for masses below 350~GeV with $\lambda_4=\lambda_5$ as large as 10 \cite{Burgess:2009wm}. Existing LHC searches that can be used to compare the MW model with the two jets channel are: ATLAS-8 TeV \cite{Aad:2014aqa}, ATLAS-13 TeV \cite{ATLAS:2016lvi},  CMS-8 TeV \cite{Khachatryan:2016ecr}, CMS-13 TeV  \cite{Khachatryan:2015dcf}. For example, ATLAS used its 8 TeV data to conclude that color octet scalars are ruled out below 2.8~TeV \cite{Aad:2014aqa}, but as explained above, their benchmark color octet model, that of Ref.~\cite{Han:2010rf}, results in cross-sections many orders of magnitude larger than the MW model.

\begin{figure}[thb]
\includegraphics[width=.45\textwidth]{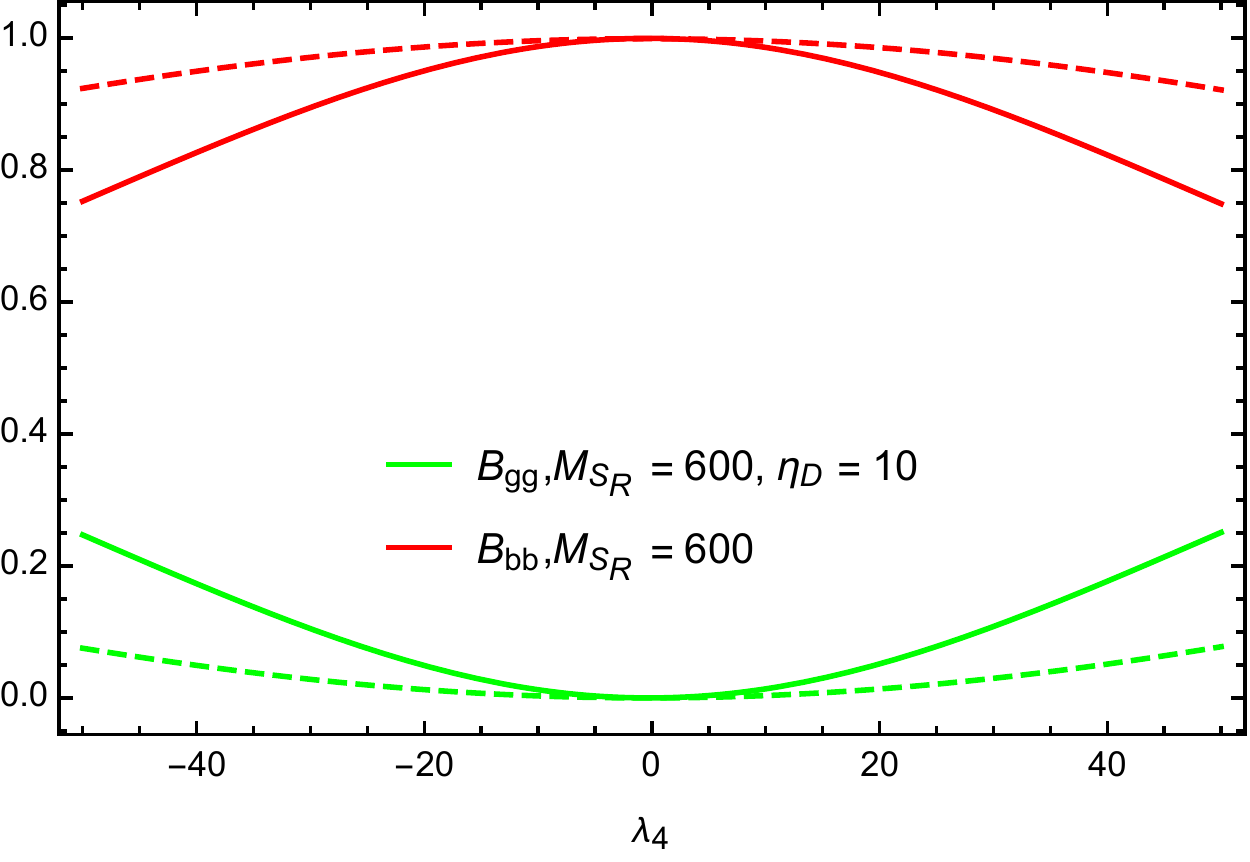}
\includegraphics[width=.45\textwidth]{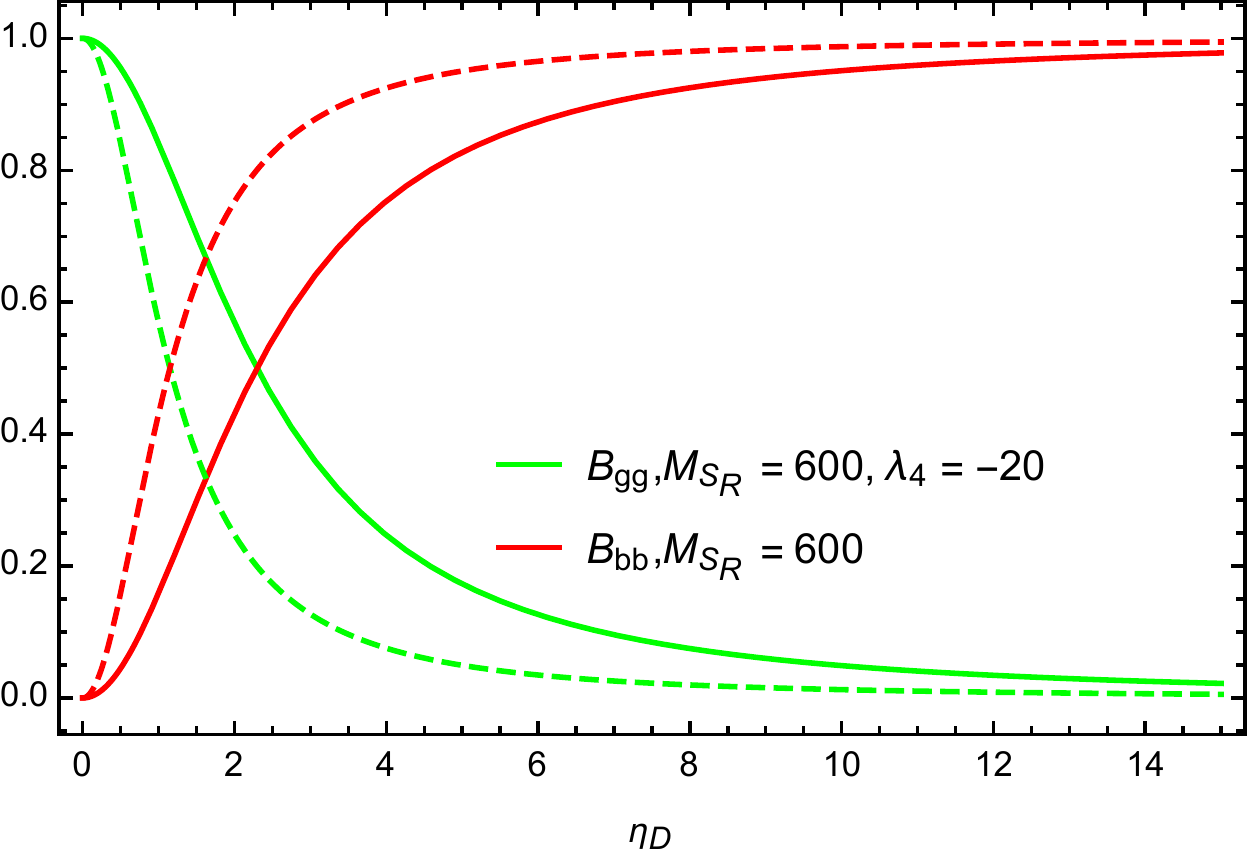}
\caption{Branching ratio for the dominant decay modes of $S_R$ with the parameters  $\eta_U=0$, $\lambda_2 = -5$. The solid lines are for $M_{S_R} = 600$~GeV and the dashed lines are for $M_{S_R} = 1200$~GeV. The higher (red) curves are for $B(S_R\to b\bar{b})$ and the lower (green) curves for $B(S_R \to gg)$. The panel on the left is for the case $\eta_D=10$ whereas the panel on the right is for $\lambda_4 = -20$.
\label{br-case1}}
\end{figure}

The dijet decay mode is dominant when one of the two neutral color octet resonances $S_R$ or $S_I$ is kinematically suppressed from decaying to the charged one through processes of the form $S_{R,I}\to S^\pm W^\mp j$, and its decay into top quark pairs is suppressed due to a small parameter $\eta_U$ (we will assume the mass is above threshold for top-pair production). Under these conditions the dominant decay rate is into two jets. We illustrate this  scenario with the choices: $\eta_U=0$, which makes $B(S_R\to t\bar{t}) \to 0 $; and $\lambda_2 = -5$, which places the mass, $M_{S_I}=M_{S^\pm}$ above $M_{S_R}$. The decay of $S_R$ in this case is almost 100\% into two jets, with the dominant channels being $b\bar{b}$ and $gg$ as illustrated in Figure~\ref{br-case1}.
The left panel on Figure~\ref{br-case1} corresponds to $\eta_D=10$ where the resonance couples most strongly to $b\bar{b}$ pairs. This coupling is weak enough that the resonance width does not exceed $\Gamma/M \lsim 10^{-3}$, a very narrow resonance which we compare to the experimental curve that assumes a width set at the detector resolution. This figure illustrates the dependence on $\lambda_4$, showing it is almost symmetric but with a slight enhancement (from constructive interference between quark and scalar loops) for negative values of $\lambda_4$. 

The panel on the right in Figure~\ref{br-case1} explores the dependence on $\eta_D$ for a fixed value $\lambda_4 = -20$. We see how the dominant decay mode turns from $gg$ to $b\bar{b}$ for values of $\eta_D$ in the $2-4$ range. In this case $\Gamma/M$ can reach the percent level for $\eta_D \gsim 30$.

\begin{figure}[h]
\includegraphics[scale=0.57]{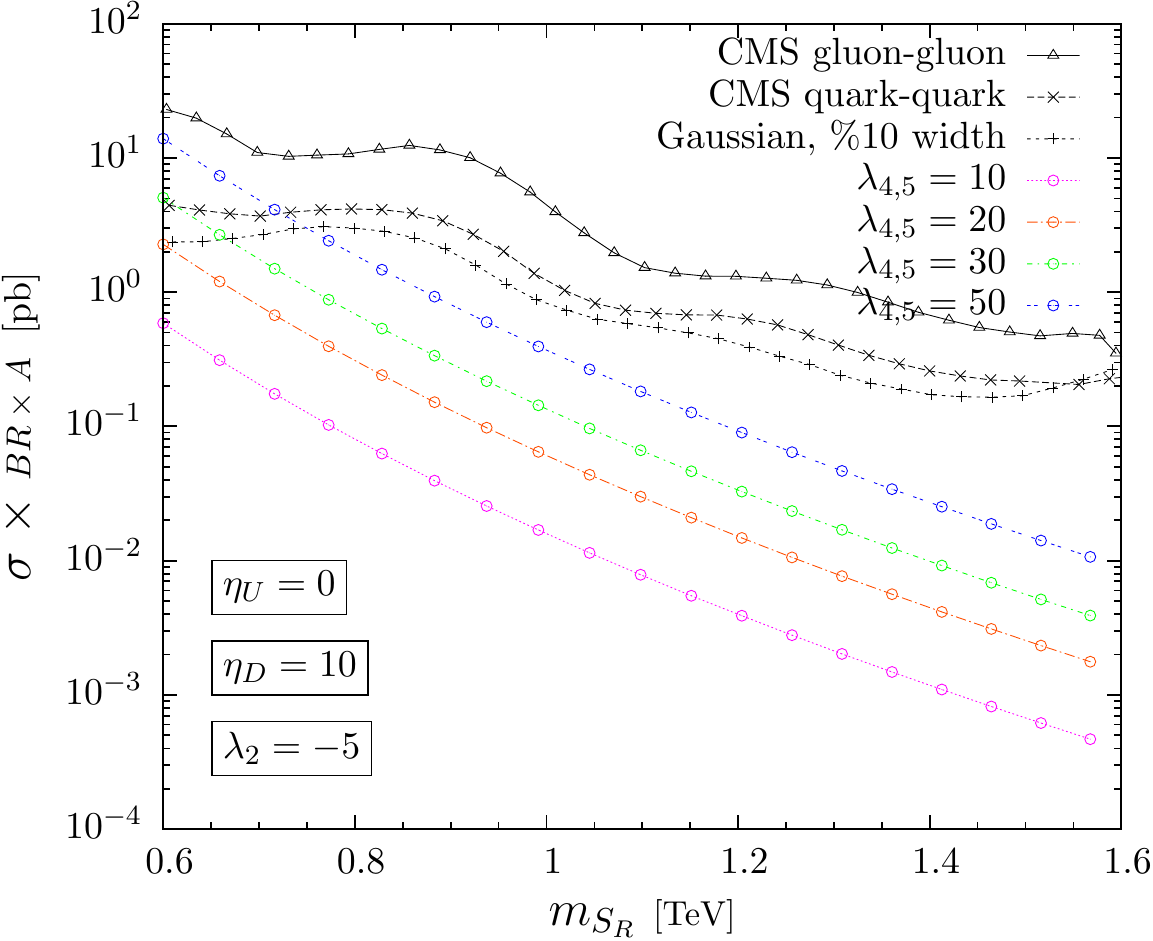} \hspace{0.5cm}
\includegraphics[scale=0.57]{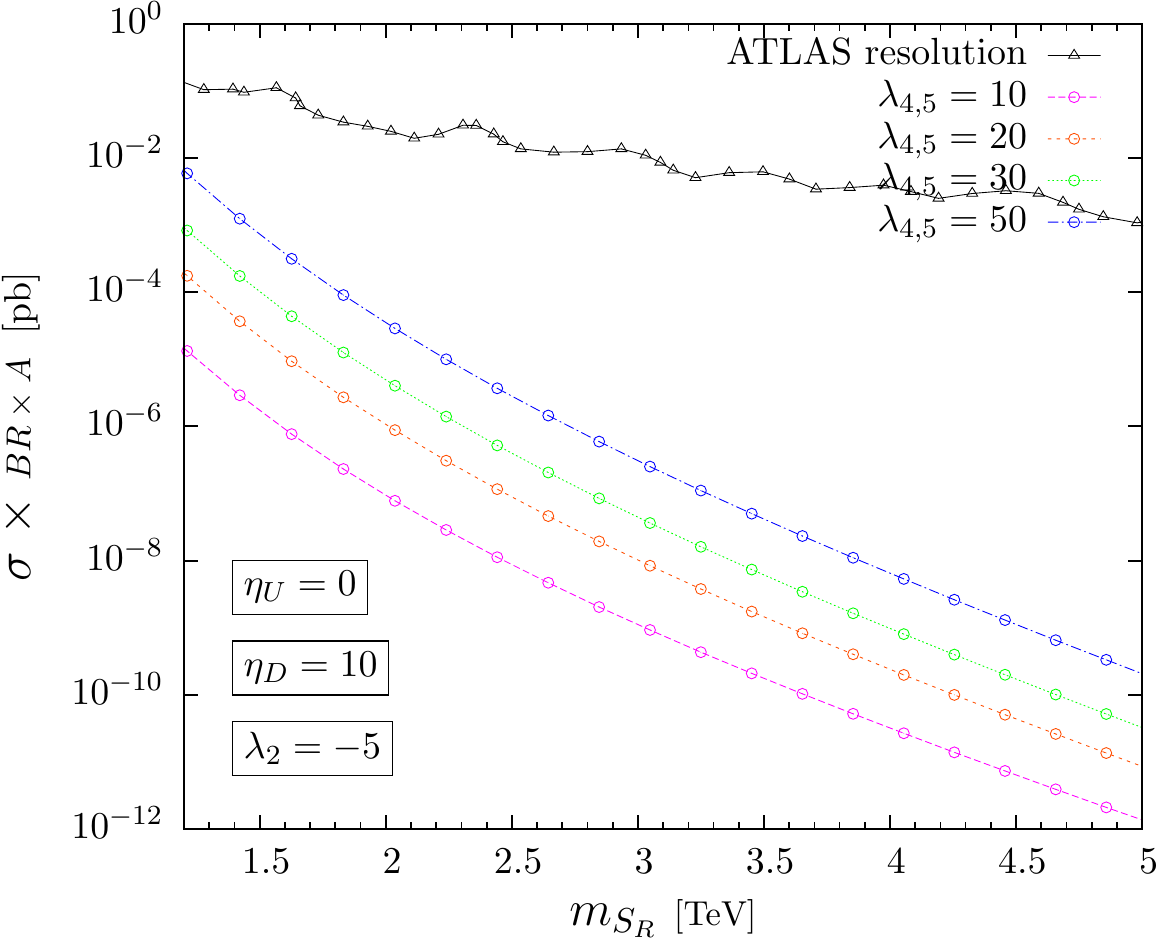} \vskip 0.3cm
\includegraphics[scale=0.57]{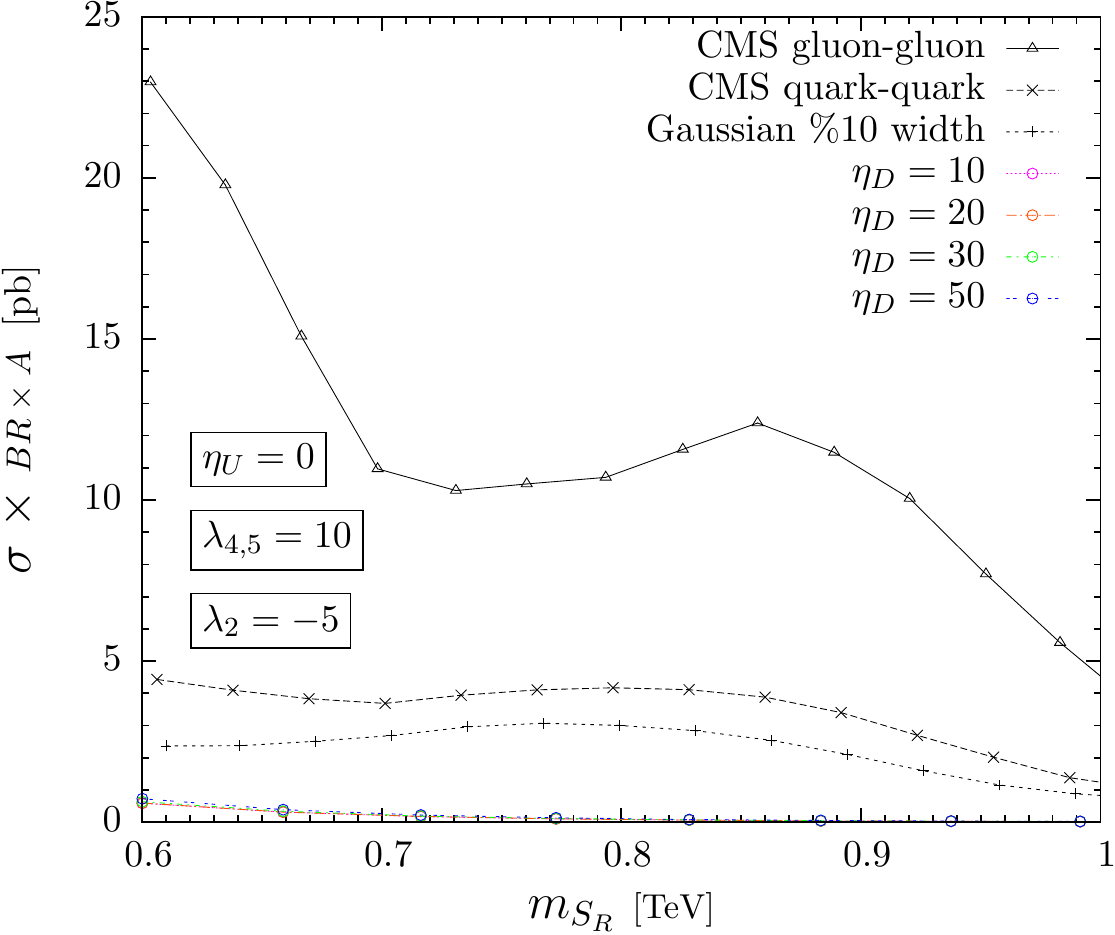} \hspace{0.7cm}
\includegraphics[scale=0.57]{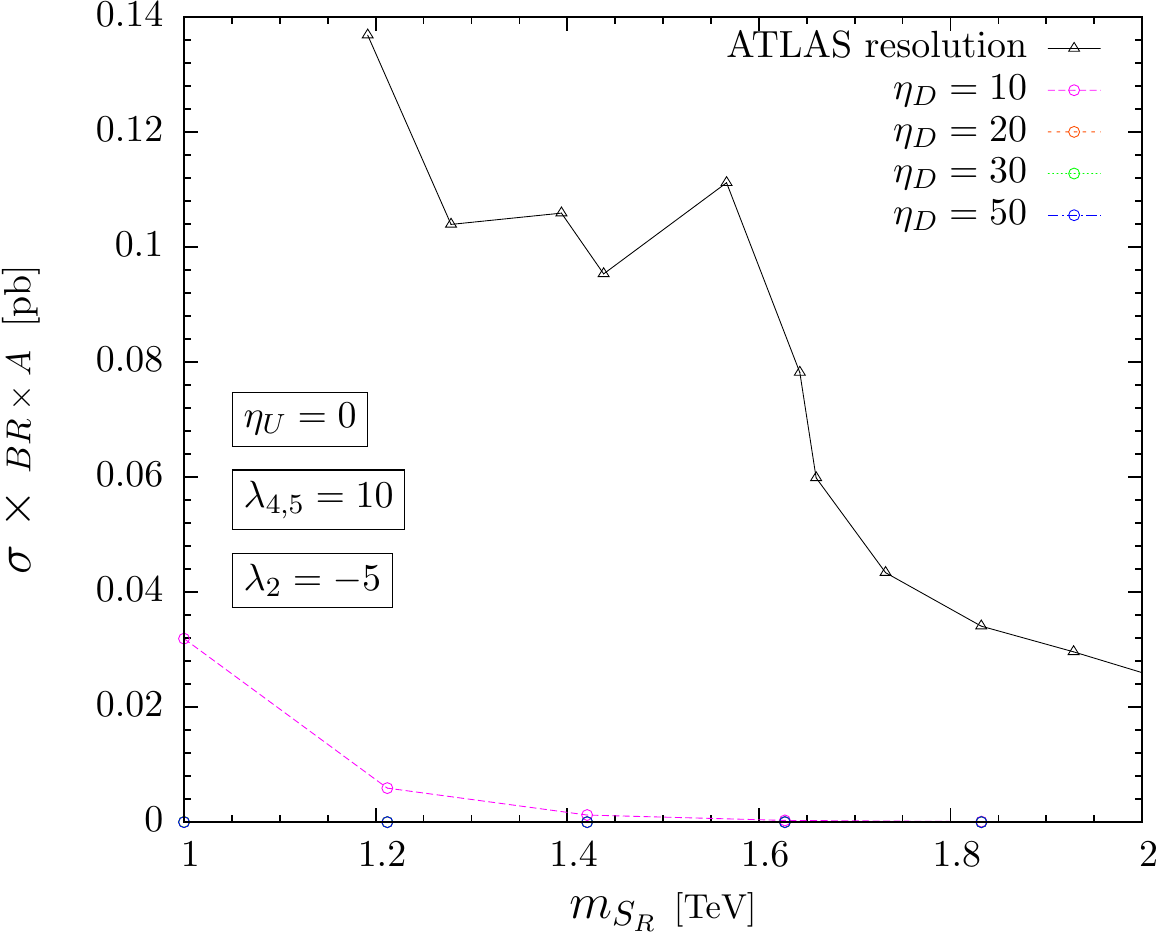}
\caption{The  panels on the left compare  production of octet scalars ($S_R$) and their decays into 2 jets in the low mass region with CMS 13TeV preliminary results \cite{CMS:2016wpz}. The panels on the right correspond to the high mass region and they are compared also with the ATLAS 13TeV preliminary results \cite{ATLAS:2016lvi}.}
\label{fig:2jets}
\end{figure}

The most restrictive data available for this channel comes from 13 TeV and we use it to extract bounds in Figure~\ref{fig:2jets}. 
The figure on the upper left panel, covers a resonance mass up to 1.6~TeV, and is best explored by CMS. For our comparison, we have used the same acceptance cuts as CMS, namely $p_{Tj}>30$~GeV, $|\eta_j|<2.5$, $|\Delta\eta_{jj}|<1.3$ and we show their results for resonances produced by (light) $q\bar{q}$ and $gg$ with 12.9 fb$^{-1}$ at 13~TeV \cite{CMS:2016wpz}. We first fix $\eta_D=10$ and present model results for different values of $\lambda_4$. Figure~\ref{br-case1} suggests that this type of resonance would be dominantly produced by a $b\bar{b}$ initial state so that it doesn't match either of the two cases explicitly considered by CMS. For example, with $\lambda_4=10$ and $M_{S_R}=600$GeV, production from $b\bar b$ accounts for 98.7\% of the cross-section, with only  about 1.1\% originating in gluon fusion. For $\lambda_4=30$, the contribution from gluon fusion has increased to about 10\%.

From this comparison we conclude that for values of $\lambda_4 \lsim 13$, its perturbative unitarity bound \cite{He:2013tla}, the cross-sections are too small for the LHC to place any meaningful limit. The figure also shows that the resonance mass is constrained to be larger than about $600\sim800$~GeV for larger values of $\lambda_4$, when we ignore the perturbative unitarity constraint. 

In the upper right panel we repeat the exercise using the results available for a higher range of resonance mass. We use in this case the ATLAS result with 15.7 fb$^{-1}$ \cite{ATLAS:2016lvi} for a hypothetical resonance with a Gaussian distribution and a width set at the detector resolution (2-3\% of the mass) and implement the ATLAS acceptance cuts $p_{Tj}>440$~GeV, sub-leading jet $p_{Tj}>60$~GeV, $|y^*|<0.6$ with $y^* = (y_3-y_4)/2$. We see that the high energy region does not yet constrain this set of parameters.

On the bottom panels we repeat the  comparisons, but this time we fix $\lambda_4=10$ and vary $\eta_D$ finding no constraints for values of $10<|\eta_D| < 50$.

\subsection{Search for resonances in top-pair production at LHC}\label{sectp}

A comparison with Tevatron cross-sections in the mass range $350-1000$~GeV found that this model could have a large enough cross-section to be relevant for $\lambda_4=\lambda_5\sim75$ which is well beyond the perturbative unitarity bound \cite{Burgess:2009wm}. In this section we use LHC data from ATLAS with 20.3 fb$^{-1}$ at 8 TeV \cite{Aad:2015fna} and from CMS with 2.6 fb$^{-1}$ at 13 TeV \cite{CMS:2016zte,CMS:2016ehh} to obtain limits from top-pair production. Once the resonance mass is above the top-pair threshold, its branching ratio into $t\bar{t}$ will be completely dominant if we suppress decays into $Wj$ or $Zj$ by choosing $\lambda_2 = -5$ as in the two-jet case and if $\eta_U$ is of order one. We perform two comparisons with data, in the first one we set $\eta_U=1$ and explore the sensitivity to $\lambda_4$ and in the second one we set $\lambda_4 = 10$ and explore the sensitivity to $\eta_U$. The $B(S\to t\bar{t})$ is at least 95\% for all of the parameters studied (with the second most dominant decay mode being two gluons), and the width of the resonance is close to 1\% of its mass for $\eta_U=1$ but grows to about 70\% by the time $\eta_U=10$ so we do not consider any larger values. The value of $\eta_D$ is set to 0 for illustration but it is irrelevant for this channel. 

 \begin{figure}[h]
\centering
\includegraphics[scale=0.57]{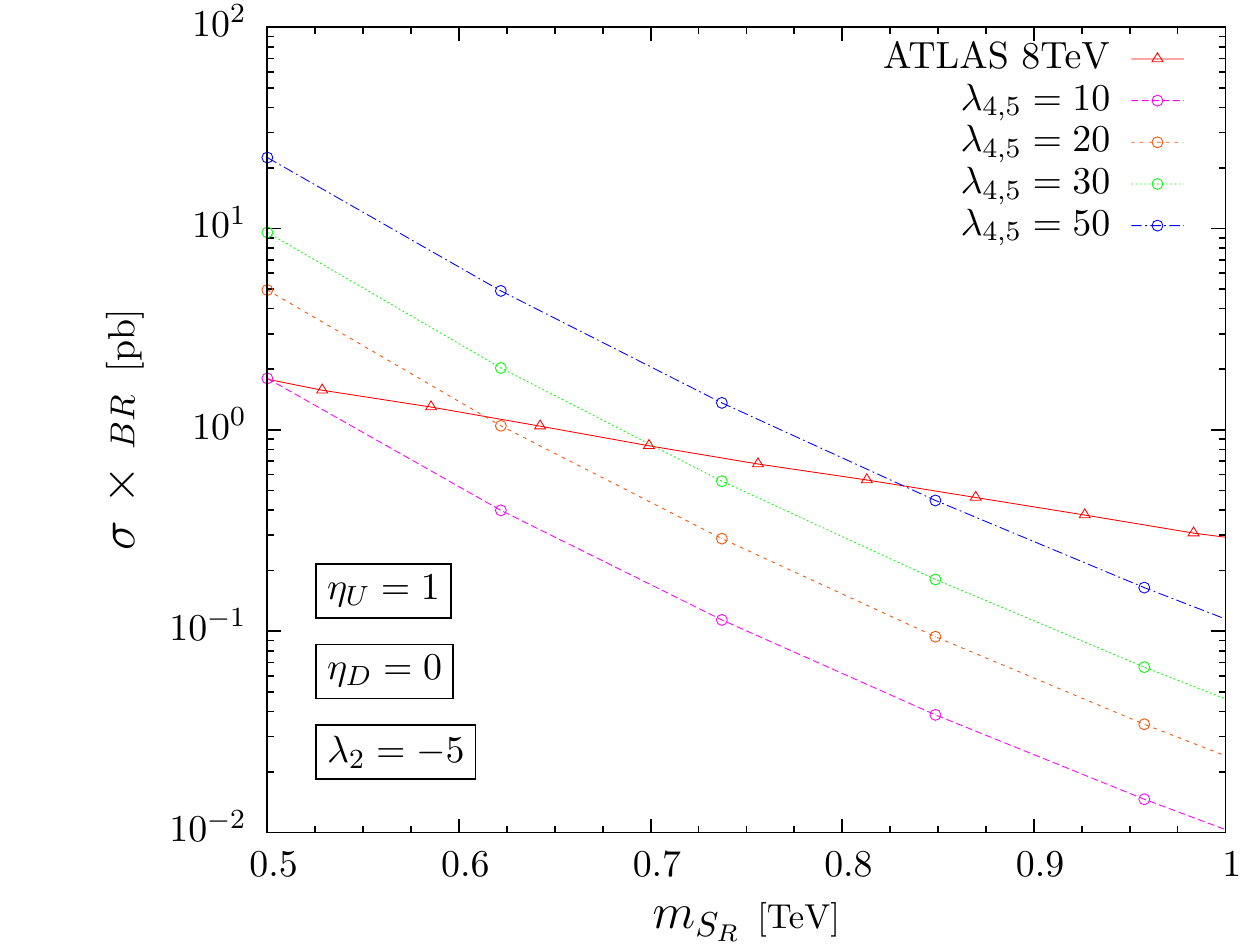}
\includegraphics[scale=0.57]{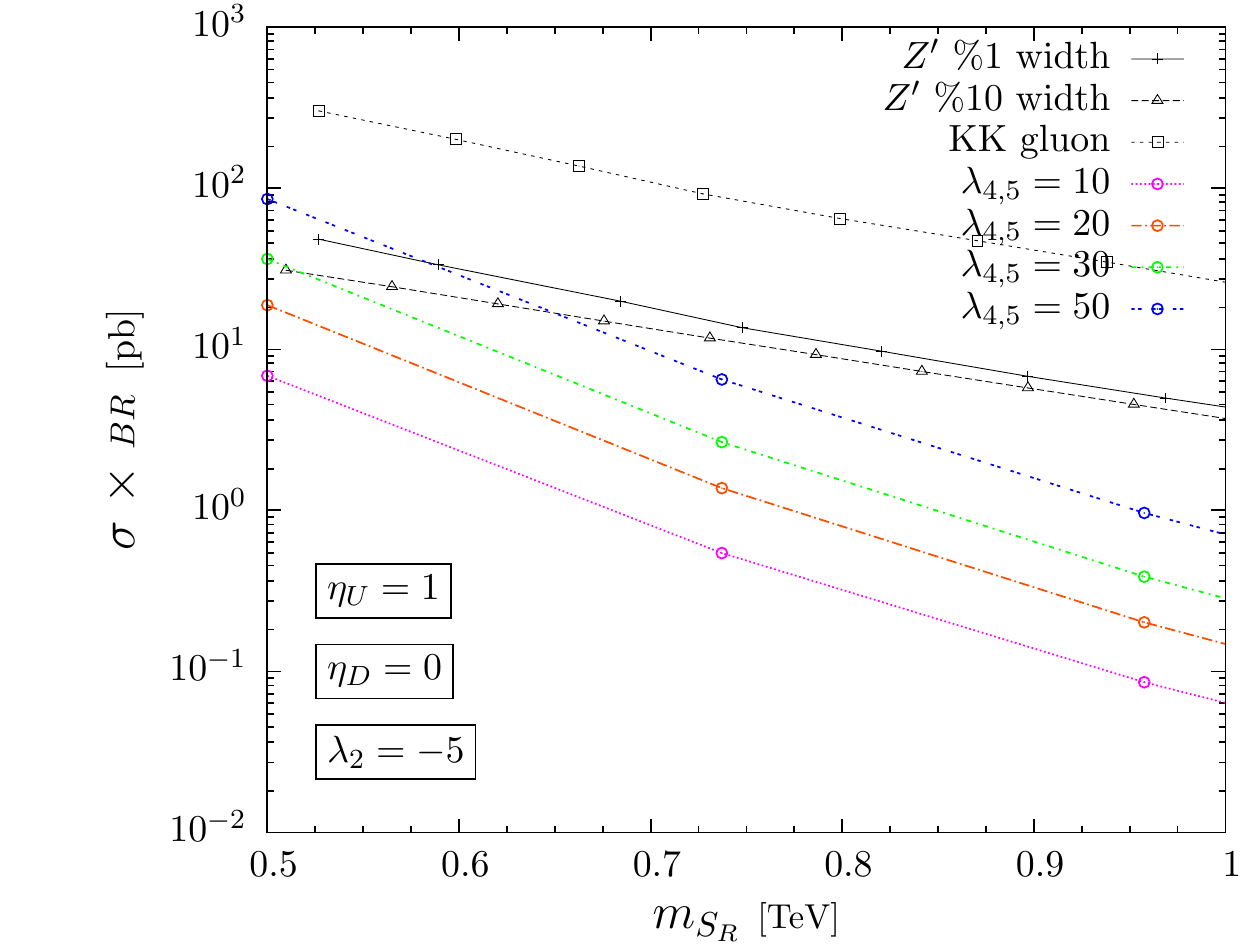} \vskip 0.3cm
\includegraphics[scale=0.57]{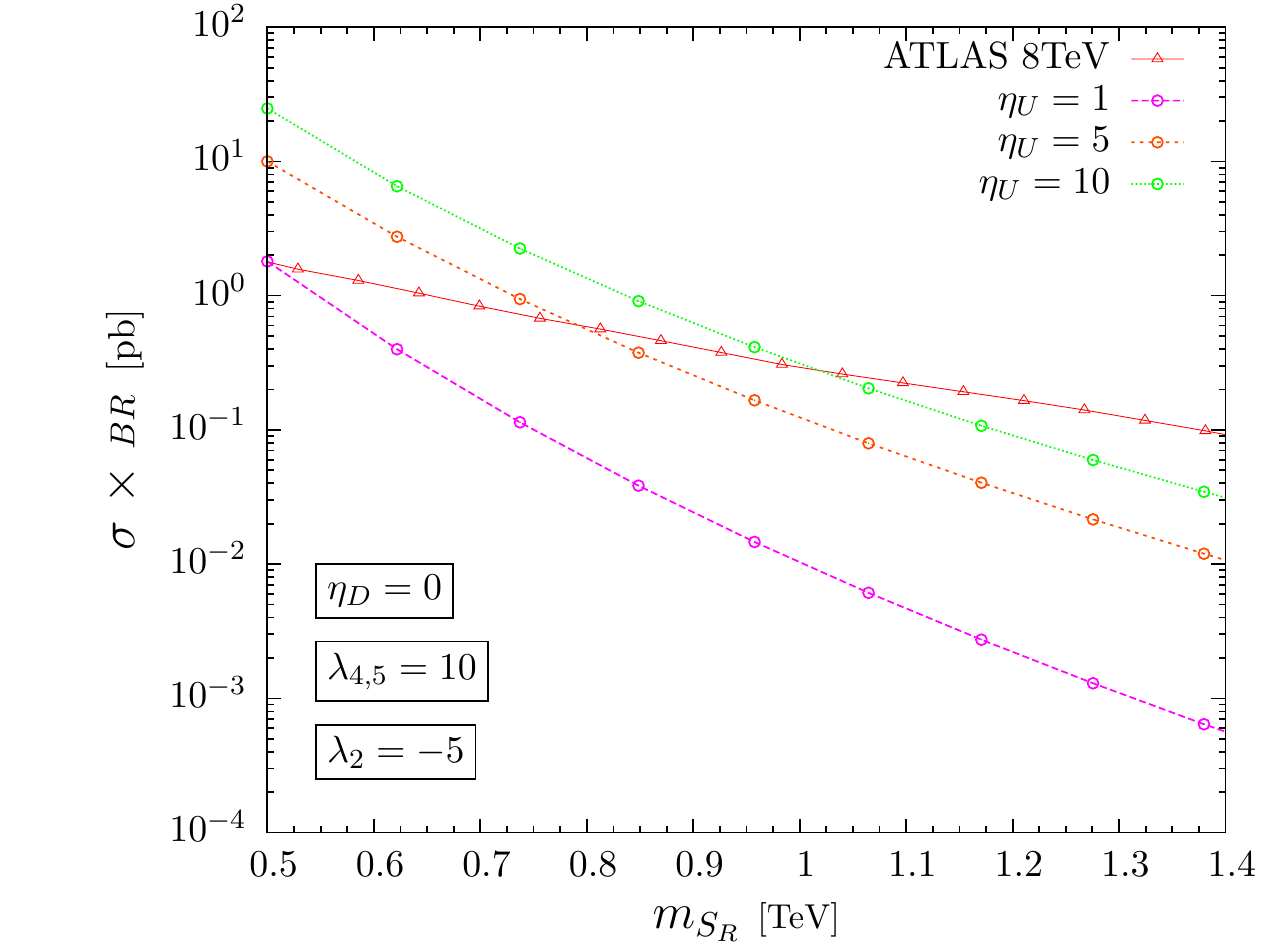}
\includegraphics[scale=0.57]{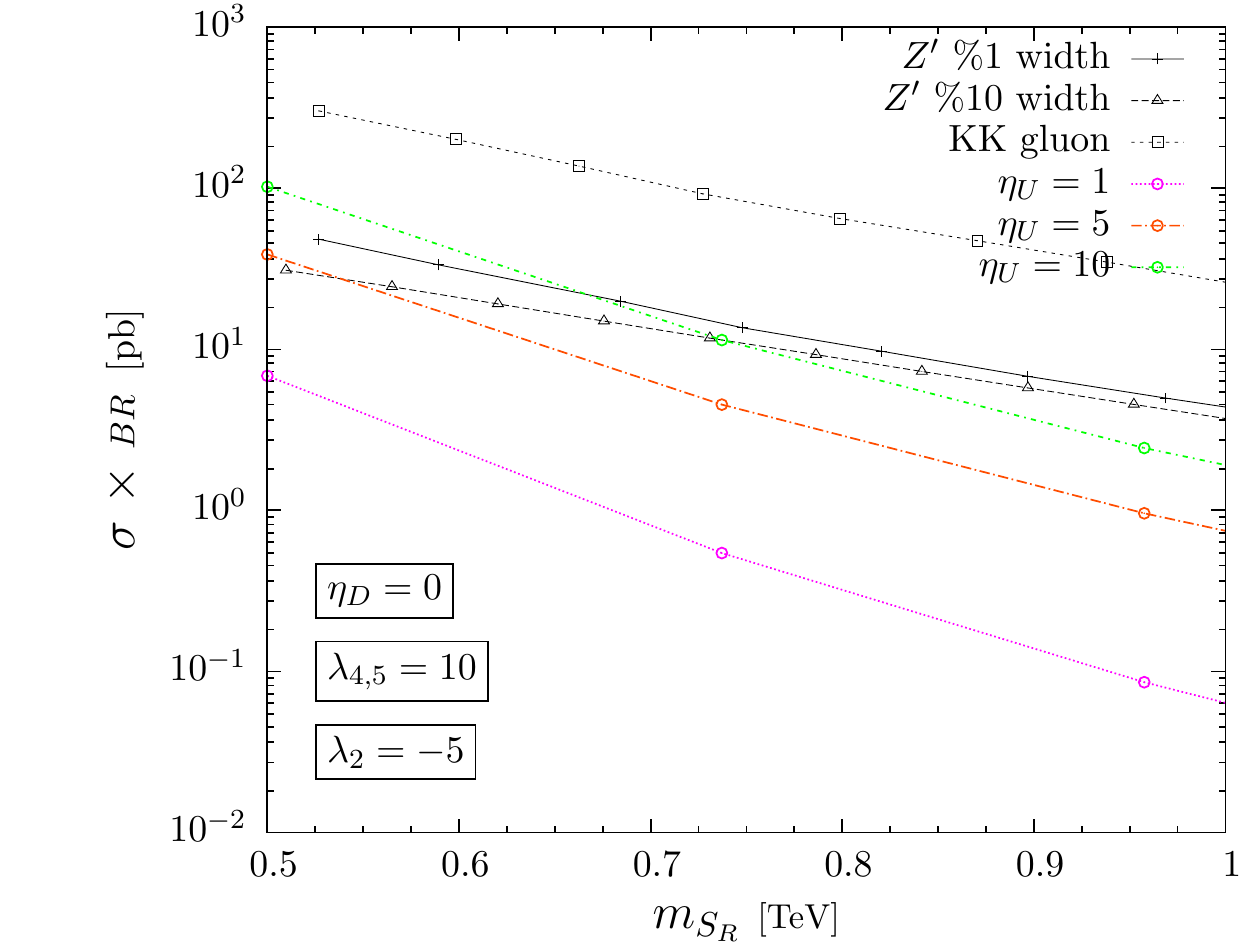}
\caption{The left figure is the production of octet scalars ($S_R$) and its decays into 2 tops at 8TeV compared with Figure 11-d of ATLAS Preliminary results \cite{Aad:2015fna} (red curve). The right figure is the same with the left but at 13TeV and it is compared with Figure 8 of CMS Preliminary results \cite{CMS:2016zte} (black curves).}
\label{fig:2tops}
\end{figure}

In the left panel we compare to the ATLAS result \cite{Aad:2015fna} for their benchmark spin 0 resonance produced by gluon fusion, which is the closest match to our discussion. Their resonance width is lower than 15\% which is adequate for $|\eta_U| \sim 5$, when the width of $S_R$ is 17\% of its mass. From the upper panel we see that $m_S \gsim 500$~GeV is obtained for values of $\lambda_4$ within its perturbative unitarity limit. The bottom panel indicates that for $\eta_U=5$ the LHC constrains $m_S \gsim 800$~GeV. For the larger value of $\eta_U=10$ the width of $S_R$ is much larger than that assumed by ATLAS, but if we still use this result, we find $m_S \gsim 1$~TeV.

On the right hand panel we have used the CMS results  \cite{CMS:2016zte} presenting three of their benchmark studies: a generic $Z'$ produced by $q\bar{q}$ with decays to $t\bar{t}$ and a width equal to 1\% or 10\% of its mass; and a KK gluon with about 94\% branching ratio to top-pairs. Since none of these scenarios closely resembles $S_R$, we can only obtain a very rough limit of $m_S \gsim 700$~GeV for large values of $\eta_U$ and no constraints for $\eta_U=1$.

\subsection{Search for resonances in dijet pairs at LHC}

\begin{figure}[h]
\centering
\includegraphics[scale=0.7]{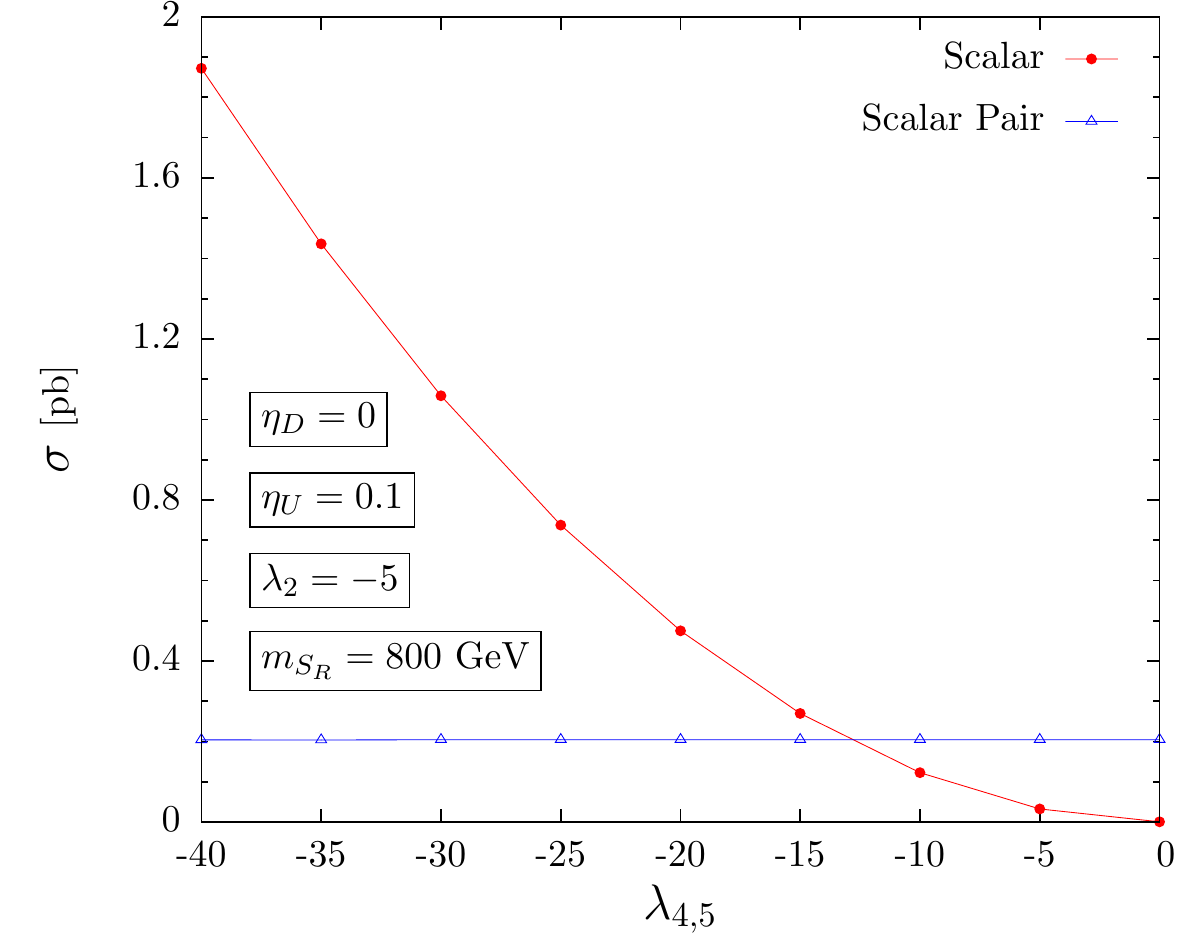}
\caption{Comparison of single $S_R$ production  vs  $S_RS_R $ scalar pair production cross-sections at LHC13 for representative values of the parameters.}
\label{SvsSS}
\end{figure}

The relevant mechanism for this channel corresponds to tree-level pair production of the neutral scalars through gluon fusion \cite{Dobrescu:2007yp}, followed by decays into light quarks or gluons. The production cross-section depends only on the mass of the scalars and the different decay channels can be selected as before. NLO calculations for the pair production of scalar color octets exist \cite{GoncalvesNetto:2012nt,Chivukula:2013hga,Degrande:2014sta} but we restrict ourselves to a LO calculation here, including a K-factor from \cite{GoncalvesNetto:2012nt}. Scalar pair production becomes dominant over single scalar production when the one-loop diagrams responsible for the latter are suppressed. This can happen for low enough values of $\eta_U$, $\lambda_4$ and $\eta_D$ as we illustrate in Figure~\ref{SvsSS}.

\begin{figure}[h]
\centering
\includegraphics[scale=0.57]{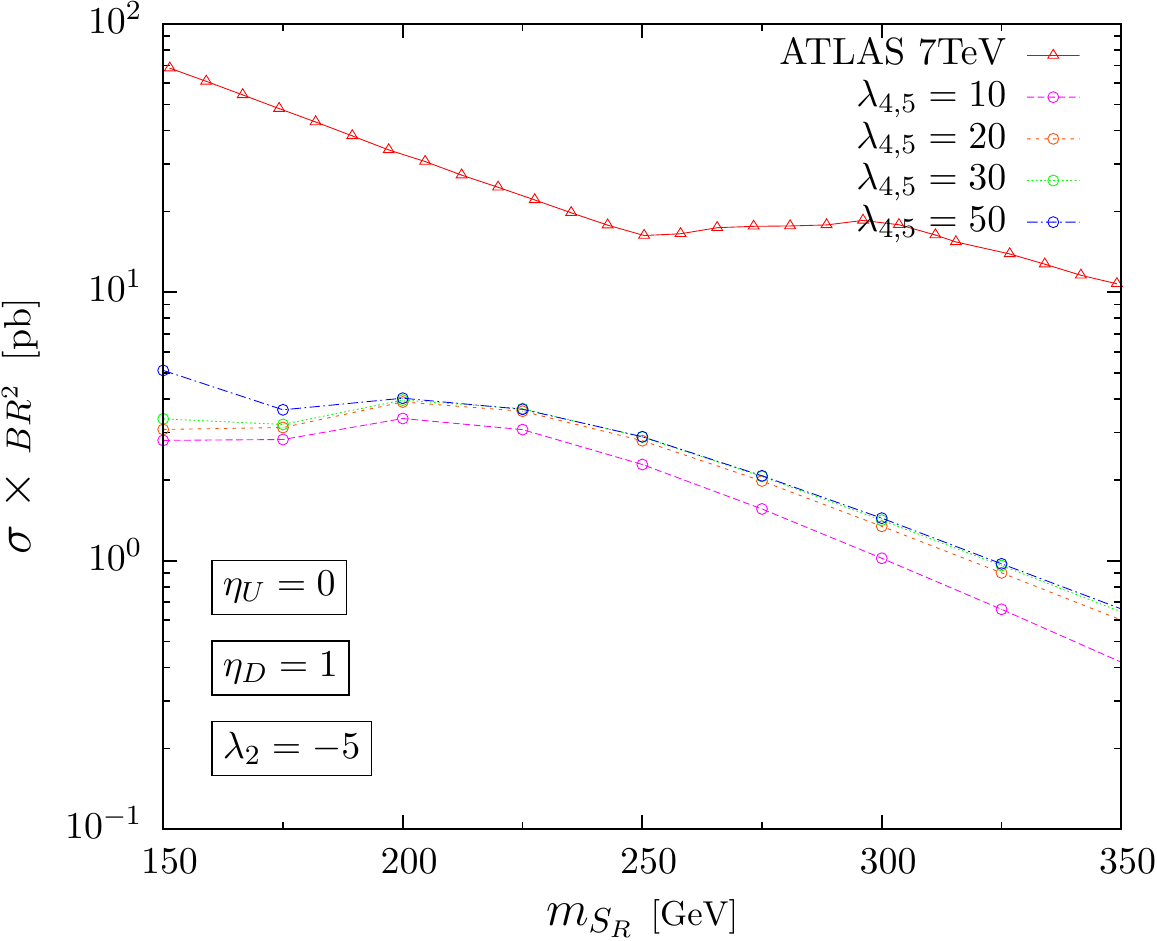} \hspace{0.5cm}
\includegraphics[scale=0.57]{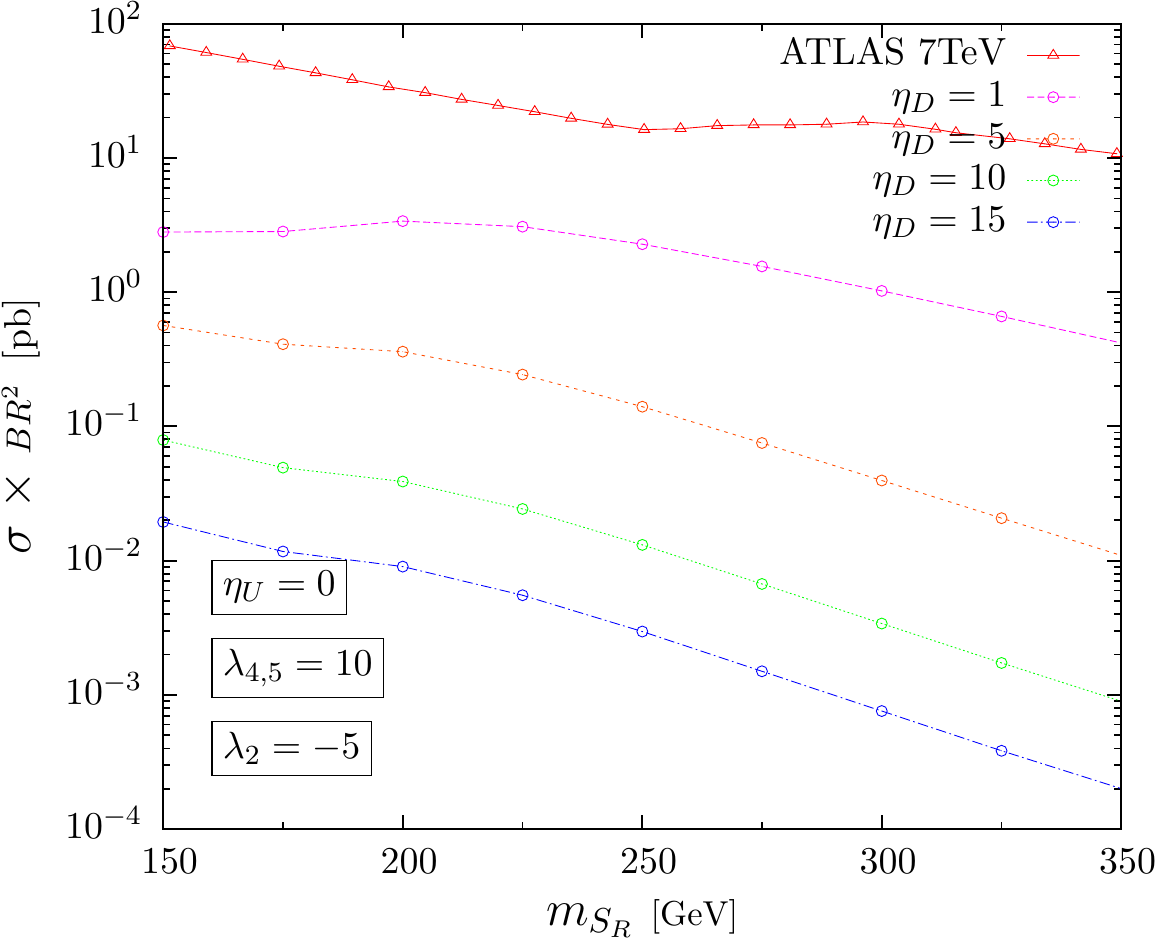} \vskip 0.3cm
\includegraphics[scale=0.57]{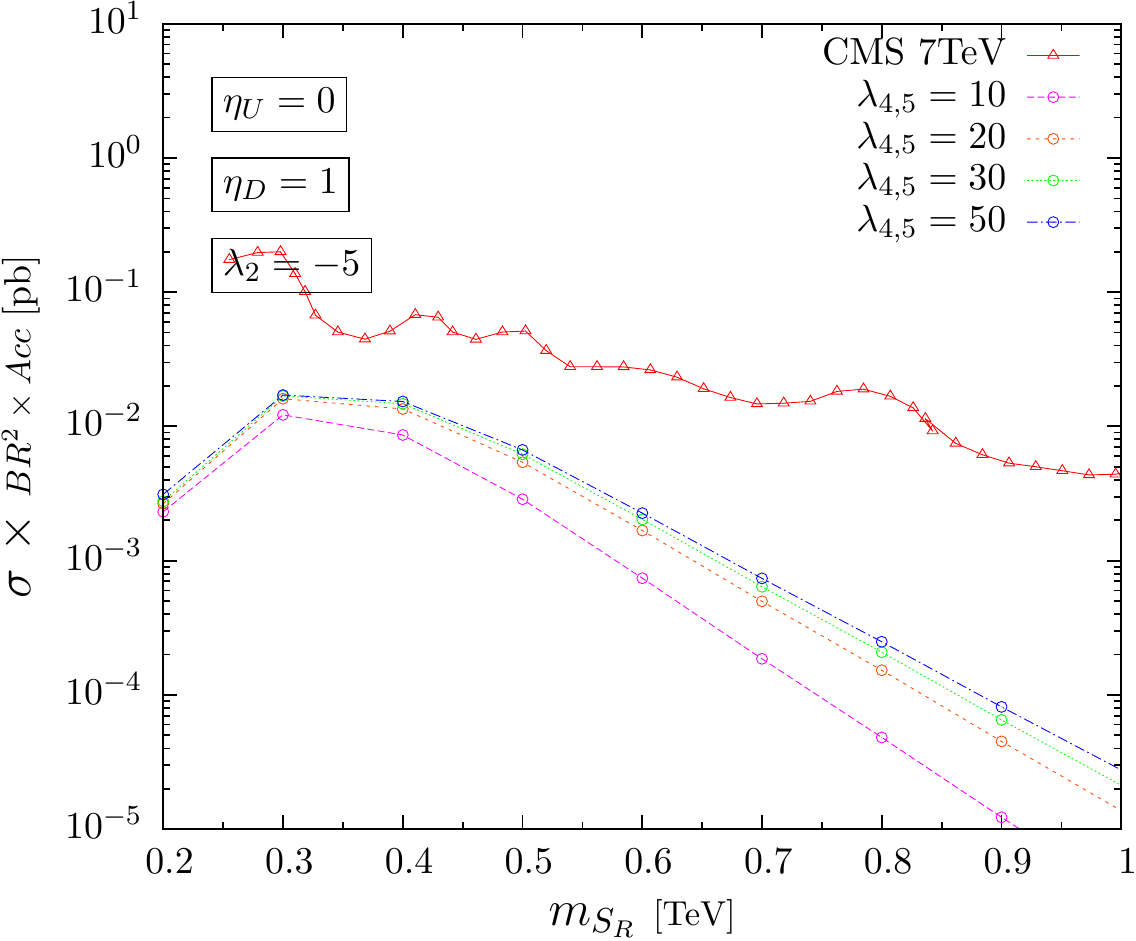} \hspace{0.5cm}
\includegraphics[scale=0.57]{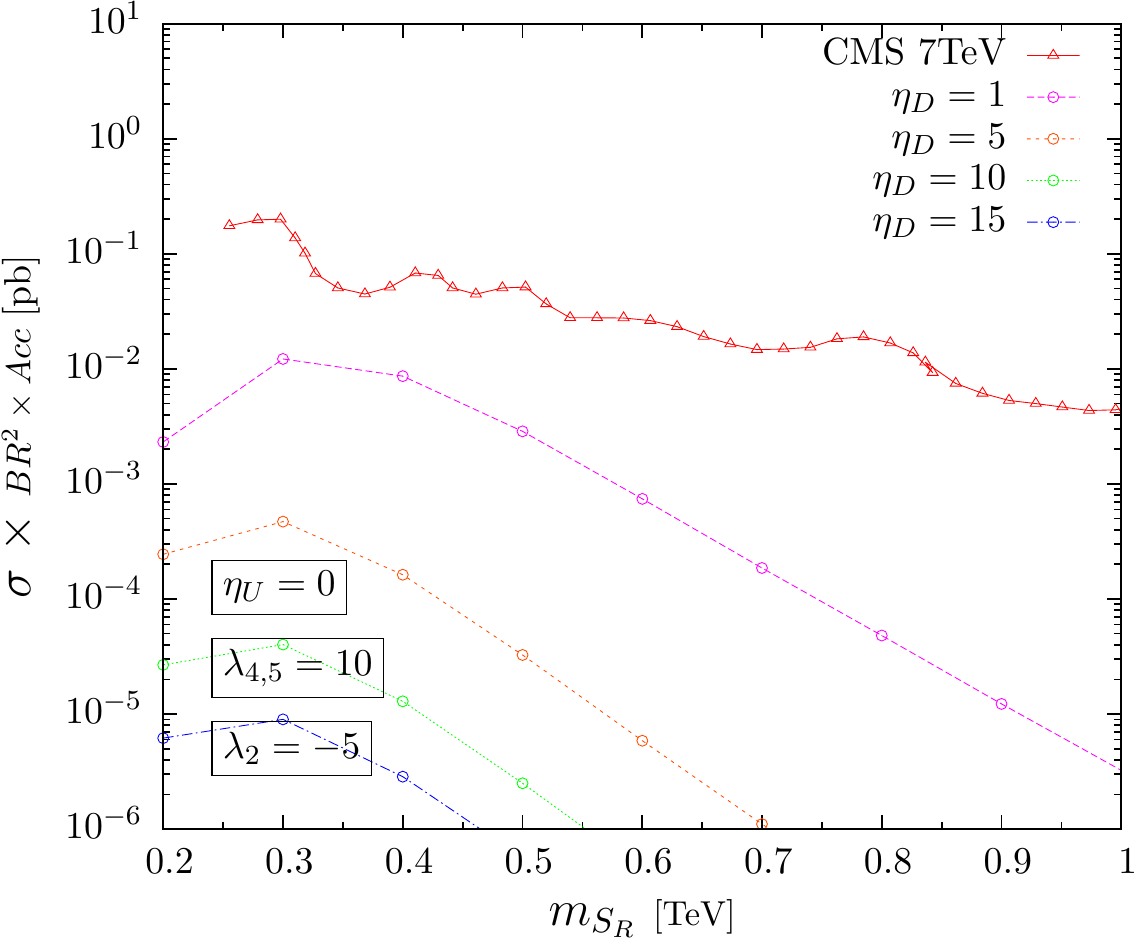} \vskip 0.3cm
\includegraphics[scale=0.57]{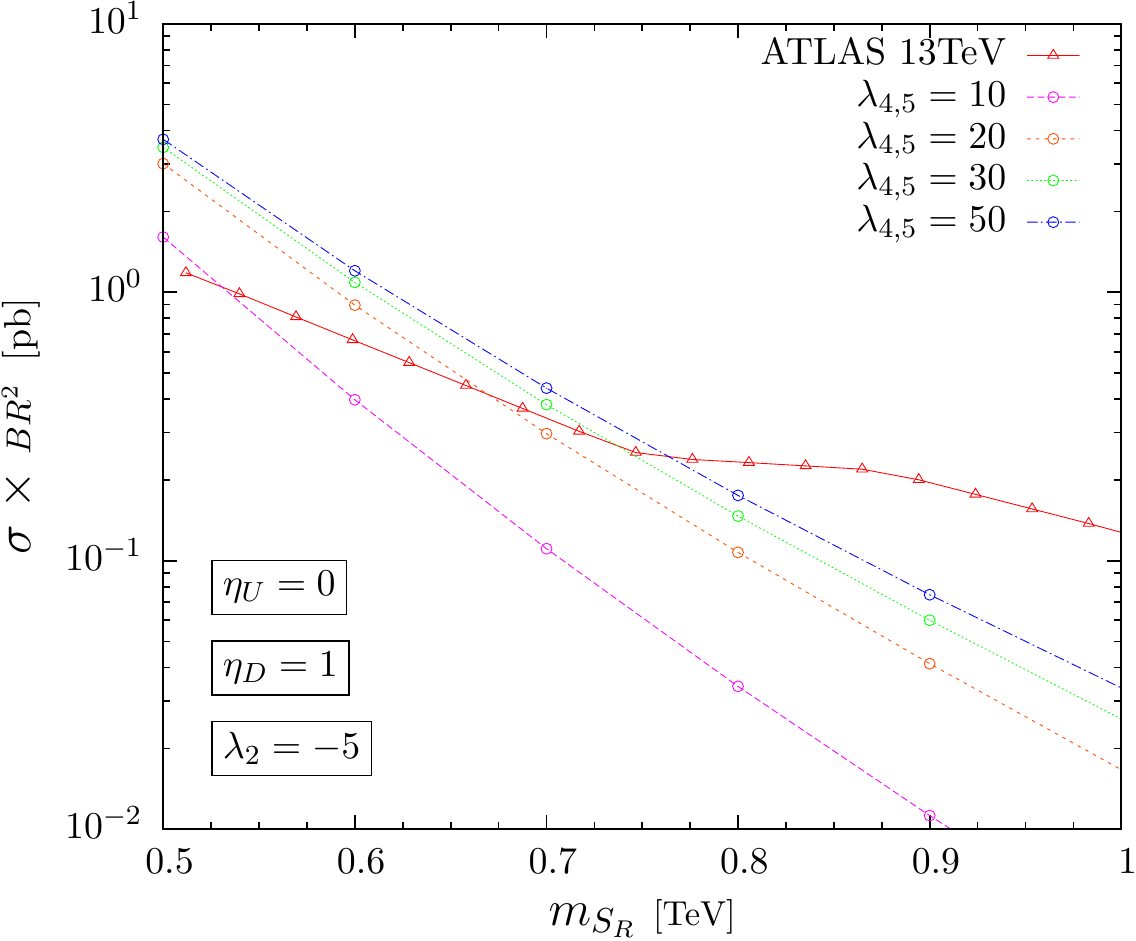} \hspace{0.5cm}
\includegraphics[scale=0.57]{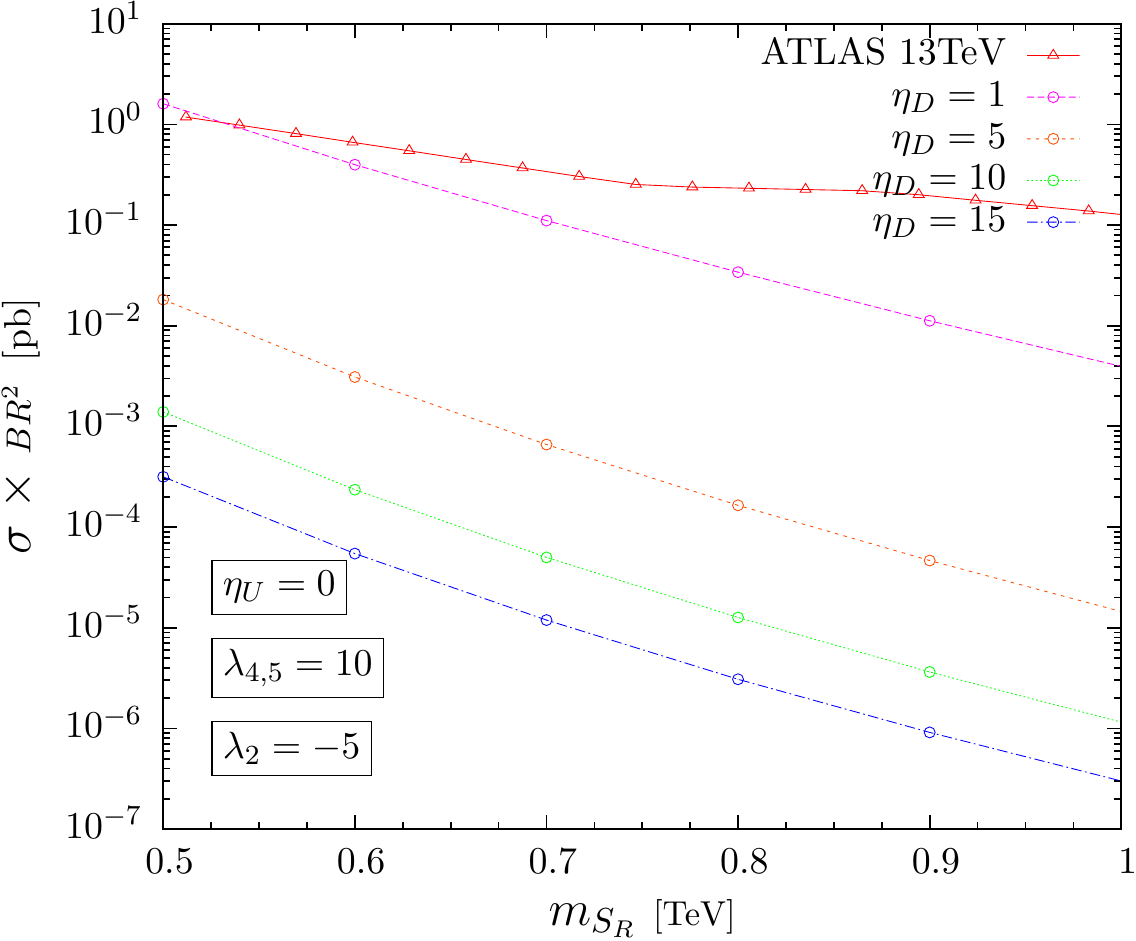}
\caption{Pair production of octet scalars ($S_R$) and their decays into 4 jets at 7~TeV (top row) compared with Figure 4 of ATLAS preliminary results \cite{ATLAS:2012ds}  (mid row) compared with Figure 3 of CMS and at 13TeV (bottom row) compared with Figure 8-b of ATLAS preliminary results \cite{ATLAS:2016sfd} (red curve).}
\label{fig:4jets}
\end{figure}

We begin with both scalars decaying to dijets, for which we suppress the decay into top-pairs with $\eta_U=0$ and the decay into $Wj$ with $\lambda_2 = -5$.  
In the upper panels of Figure \ref{fig:4jets} we compare the model to published results from ATLAS  at 7~TeV which cover the low mass region, $150-350$~GeV \cite{ATLAS:2012ds} using their acceptance cuts $p_{Tj}>80$~GeV and $|\eta_j|<1.4$. For the mass region up to 1~TeV we compare in the middle panels with the CMS preliminary results at 7~TeV \cite{Chatrchyan:2013izb}, in this case with the acceptance cuts given by $p_{Tj}>110$~GeV, $|\eta_j|<2.5$ and $\Delta R_{jj}>0.7$. Finally, in the bottom panels we compare the region below 1~TeV 
with the results from ATLAS with 15.4 pb$^{-1}$ at 13~TeV  \cite{ATLAS:2016sfd} corresponding to their benchmark scenario that most closely resembles our case, pair production of colorons. 
On the left panel of Figure \ref{fig:4jets} we fix $\eta_D=1$, for which Figure \ref{br-case1} tells us the jets will be dominantly gluon jets. The figure then explores the sensitivity to $\lambda_4$, finding that resonances as light as 500~GeV are still allowed for $\lambda_4\lsim 10$, but that masses  $\lsim 750$~GeV are excluded if we allow for larger values of $\lambda_4$. In the right side panels we take $\lambda_4=10$ and vary $\eta_D$. Only the 13~TeV data constrains the model, finding in this case that resonances with $m_S\lsim 650$~GeV are excluded with $\eta_D=1$. For larger $\eta_D$ the dominant decay is into four bottom quarks, which we treat as a separate case below.

\subsection{Search for resonances in four top events}

The $t\bar{t} t\bar{t}$, $b\bar{b} b\bar{b}$ or $b\bar{b} t\bar{t}$ final states have been discussed in the literature before. They have lower cross-sections than the respective heavy-quark pair production, but suffer from a much smaller QCD background and this makes them ideal to search for new resonances. They have studied both for pair produced resonances \cite{Beck:2015cga}, as we do here, and for a new resonance produced in association with a heavy quark pair \cite{Han:2004zh}.

\begin{figure}[h]
\centering
\includegraphics[scale=0.57]{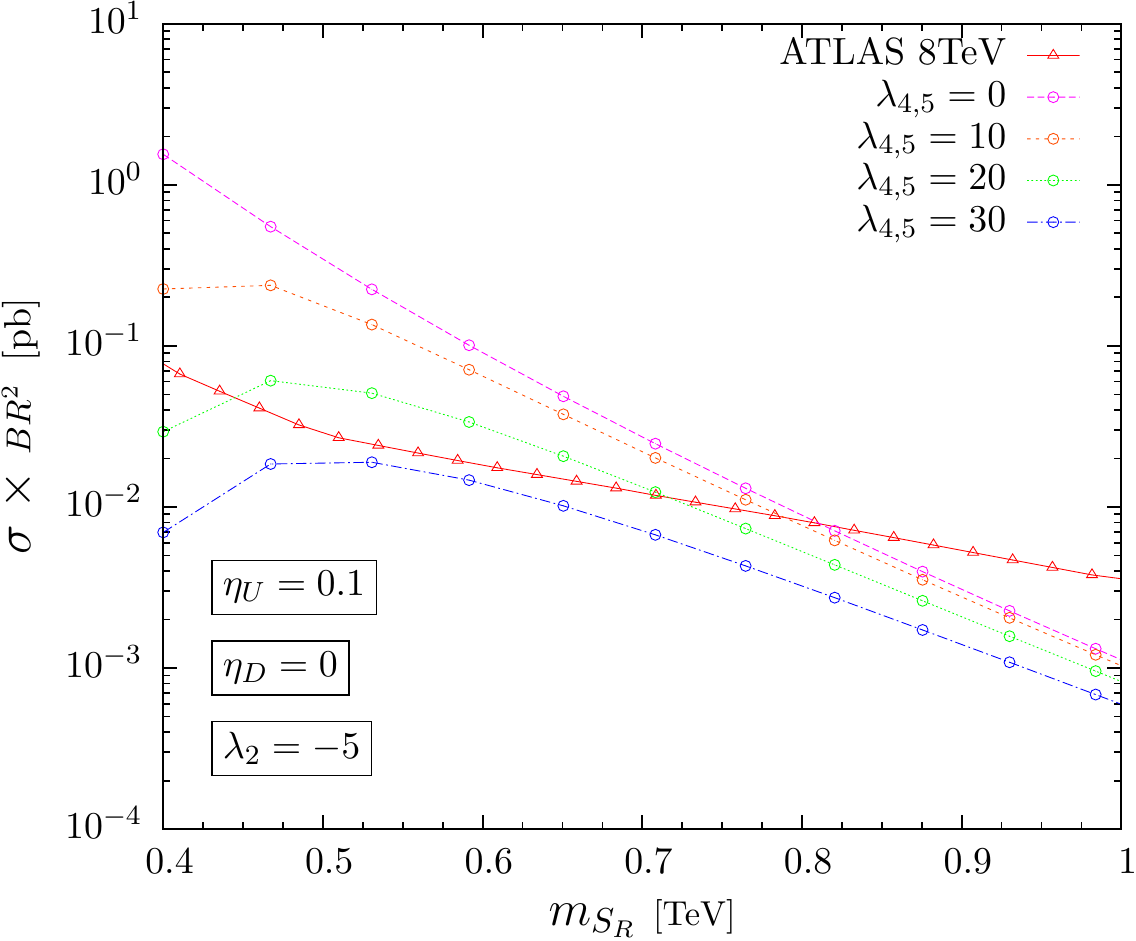} \hspace{0.5cm}
\includegraphics[scale=0.57]{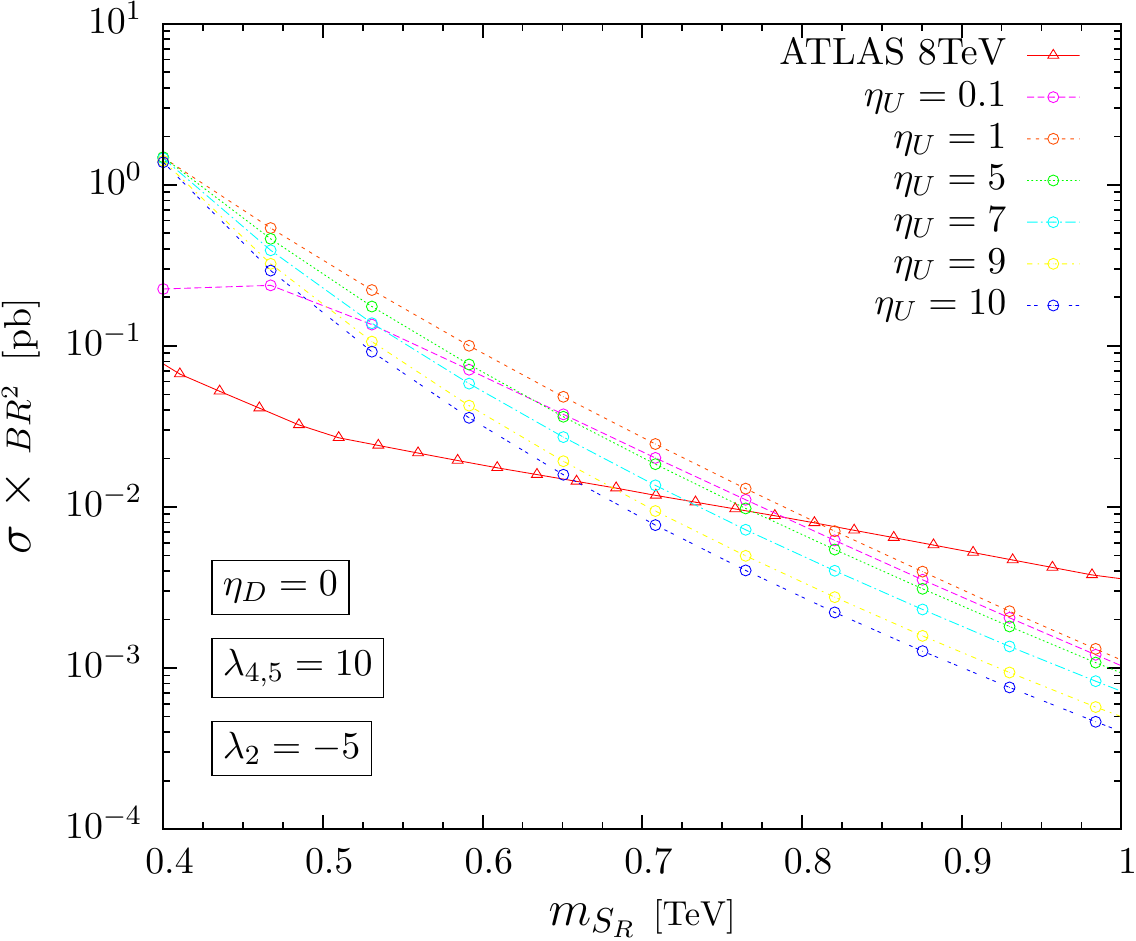} \vskip 0.3cm
\includegraphics[scale=0.57]{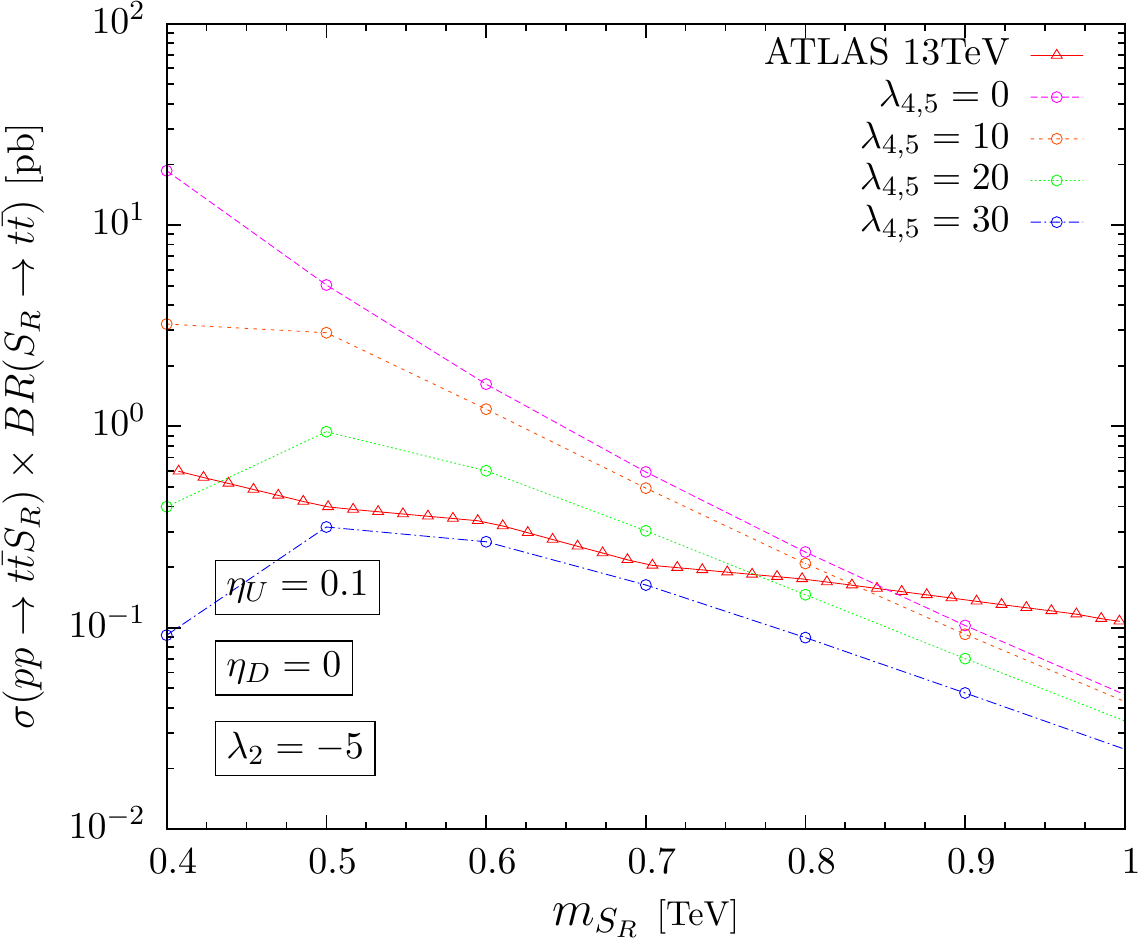} \hspace{0.5cm}
\includegraphics[scale=0.57]{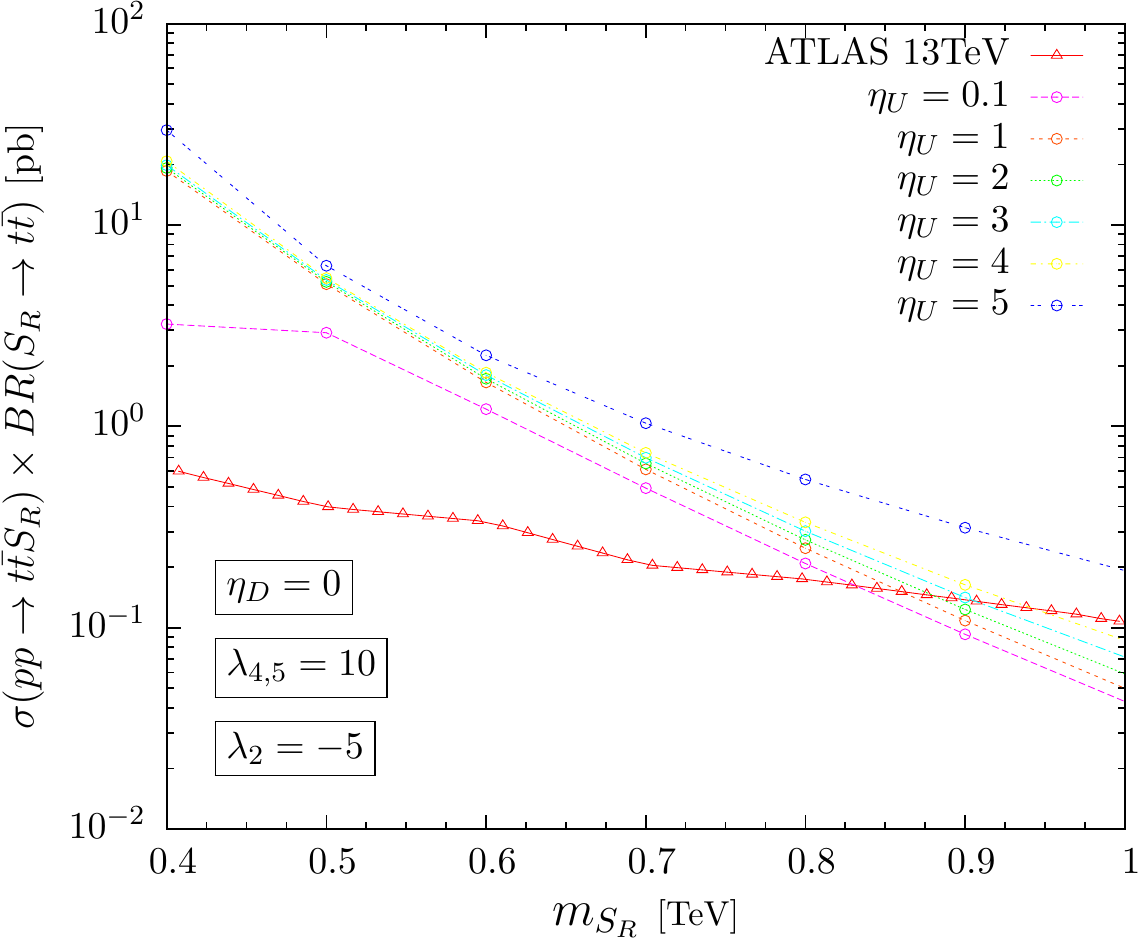}
\caption{Pair production of octet scalars ($S_R$) and their decays into 4 tops at 8TeV (upper row) compared with Figure 26 of ATLAS Preliminary results \cite{Aad:2015kqa} (red curve) and associated production of octet scalar ($S_R$) and its decay into 2 tops at 13TeV (lower row) compared with Figure 22 of \cite{ATLAS:2016btu}. }
\label{fig:4ts}
\end{figure}

As in the previous sub-section we consider pair production of scalars via their QCD couplings and adjust the model parameters so that they decay predominantly into top-pairs as we did in \ref{sectp}. We thus fix $\lambda_2=-5$, $\eta_U \gsim 0.1$ and $\eta_D=0$. For LHC data we use the ATLAS results with 20.3 fb$^{-1}$ at 8~TeV \cite{Aad:2015kqa} and their benchmark model of scalar gluon pair production with subsequent decays to two top quark anti-quark pairs. In the left side panel of Figure \ref{fig:4ts} we fix $\eta_U=0.1$ and explore the dependence on $\lambda_4$. We find that a new resonance with $m_S \lsim 800$~GeV is excluded for values of $\lambda_4$ within their perturbative unitarity limits by the 8~TeV data. In the bottom panels we also compare with 13~TeV data from associated production of $S_R$ with a top quark pair, Figure 22 of \cite{ATLAS:2016btu}. In this case a mass  as large as $m_S\sim 850$~GeV is excluded. As $\lambda_4$ increases the constraints disappear because the decay of $S_R$ into two gluons becomes dominant over $t\bar{t}$ pairs. On the right side panel we fix $\lambda_4=10$ and allow $\eta_U$ to vary. As $\eta_U$ increases further the constraint at 8~TeV gets weaker because the resonance width becomes very large, going from 16\% to  64\% of the mass for $m_S=700$~GeV as  $\eta_U$ is increased from 5 to 10, for example. For  $\eta_U \sim 5$ the exclusion extends to $m_S \lsim 1$~TeV with the 13~TeV data. 

\subsection{Search for resonances in four bottom events}

\begin{figure}[h]
\centering
\includegraphics[scale=0.57]{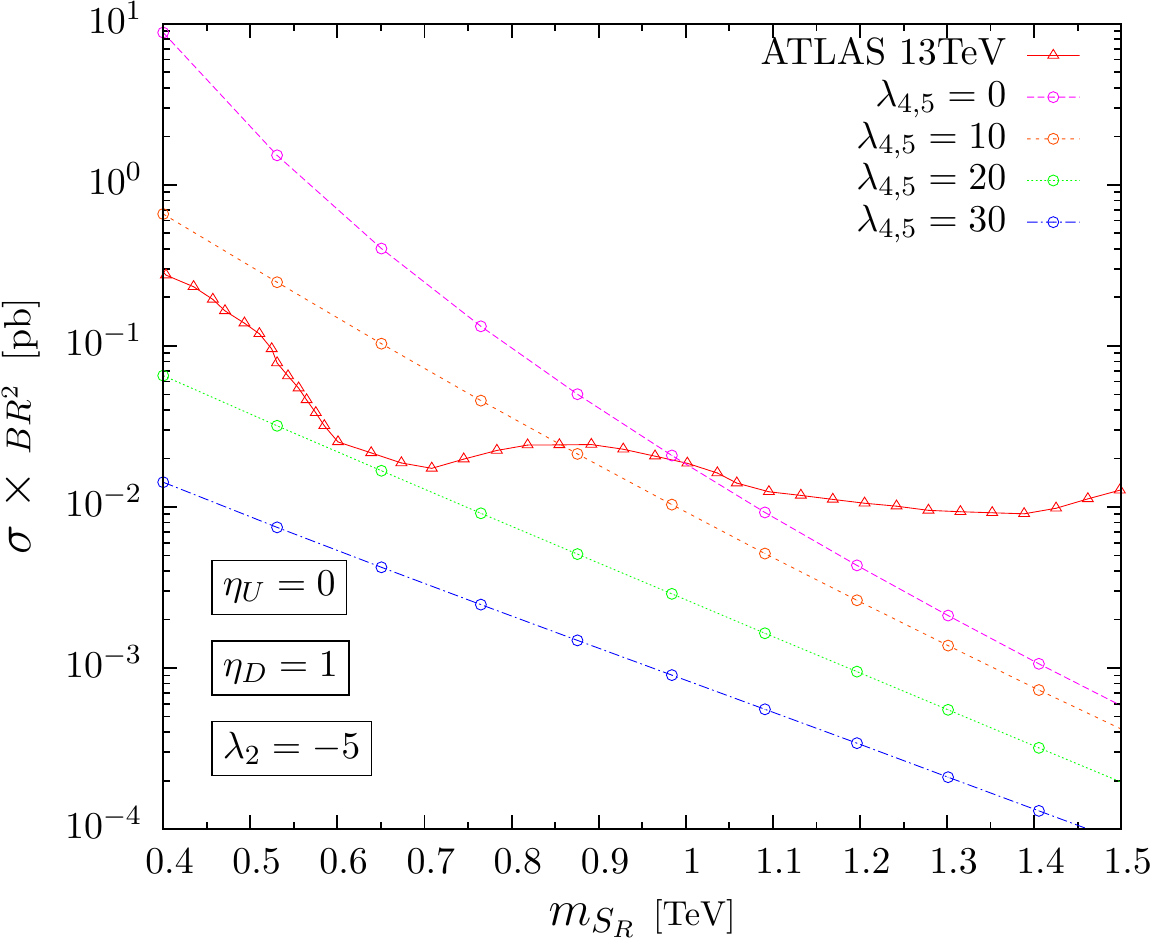} \hspace{0.5cm}
\includegraphics[scale=0.57]{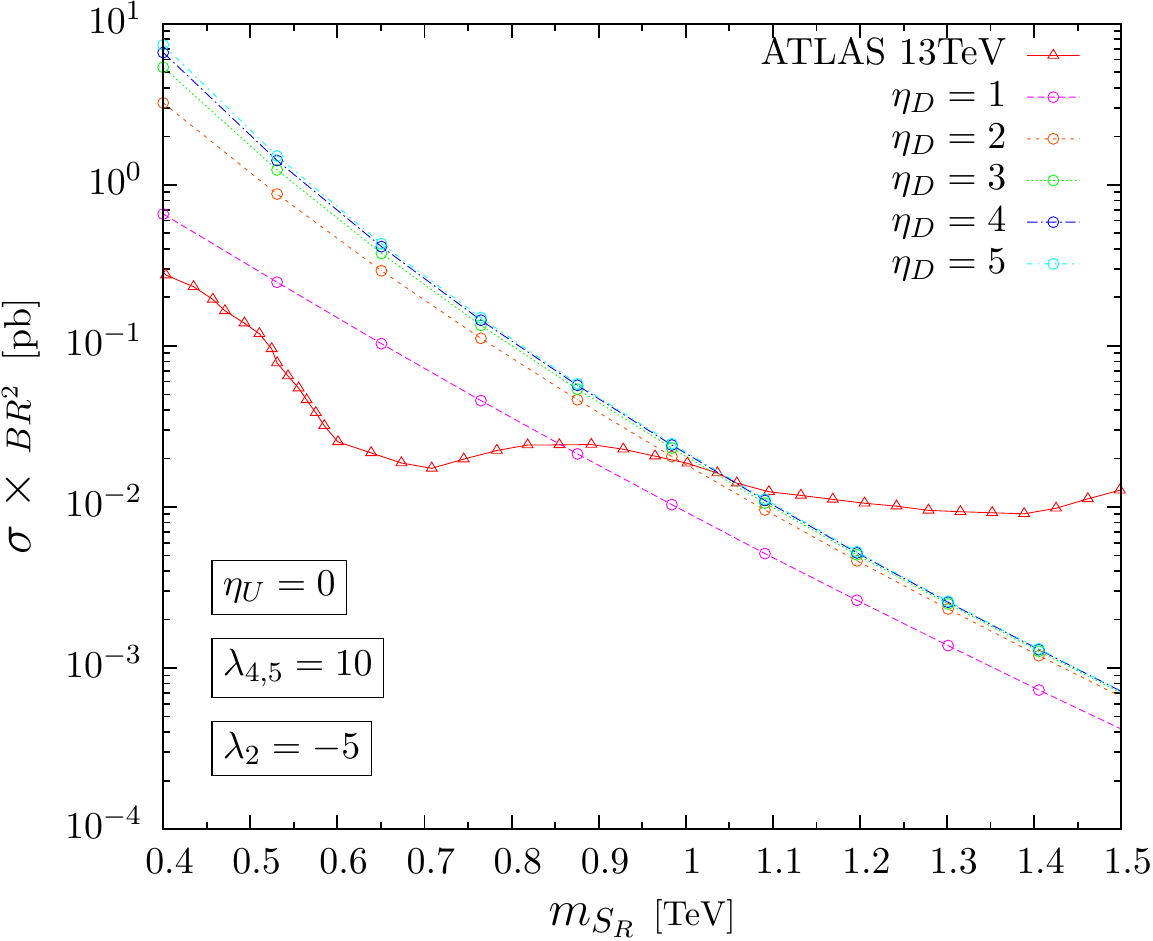} \\
	\caption{Pair production of octet scalars ($S_R$) and their decays into 4 bottoms at 13TeV compared with Figure 11 of ATLAS Preliminary results \cite{ATLAS:2016ixk} (red curve).}
	\label{fig:4bs}
\end{figure}

Finally we consider pair production of scalars via their QCD couplings and adjust the model parameters so that they decay predominantly into bottom-pairs \cite{Bai:2010dj}. We study this case by comparing it to ATLAS results from 13.3 fb$^{-1}$ at 13 TeV and their search for Higgs pair production in the $b\bar{b}b\bar{b}$ mode \cite{ATLAS:2016ixk}. In the left panel of Figure \ref{fig:4bs} we set $\eta_D=1$ and vary $\lambda_4$. For values of $|\lambda_4| \lsim 10$ this LHC study excludes $m_S \lsim 1$~TeV. At larger values of $\lambda_4$ the constraint disappears as the decay of $S_R$ into two gluons is dominant over $b\bar{b}$ pairs. On the right side panel we fix $\lambda_4=10$ and see that for $1\lsim \eta_D\lsim 5$ the region $m_S \lsim 1$~TeV is excluded. For lower values of $\eta_D$ the resonance decays to two gluons as can be seen in Figure~\ref{br-case1} and the constraint disappears. 

\subsection{Search for resonances in $t\bar{t}b\bar{b}$ events}

\begin{figure}[h]
\centering
	\includegraphics[scale=0.57]{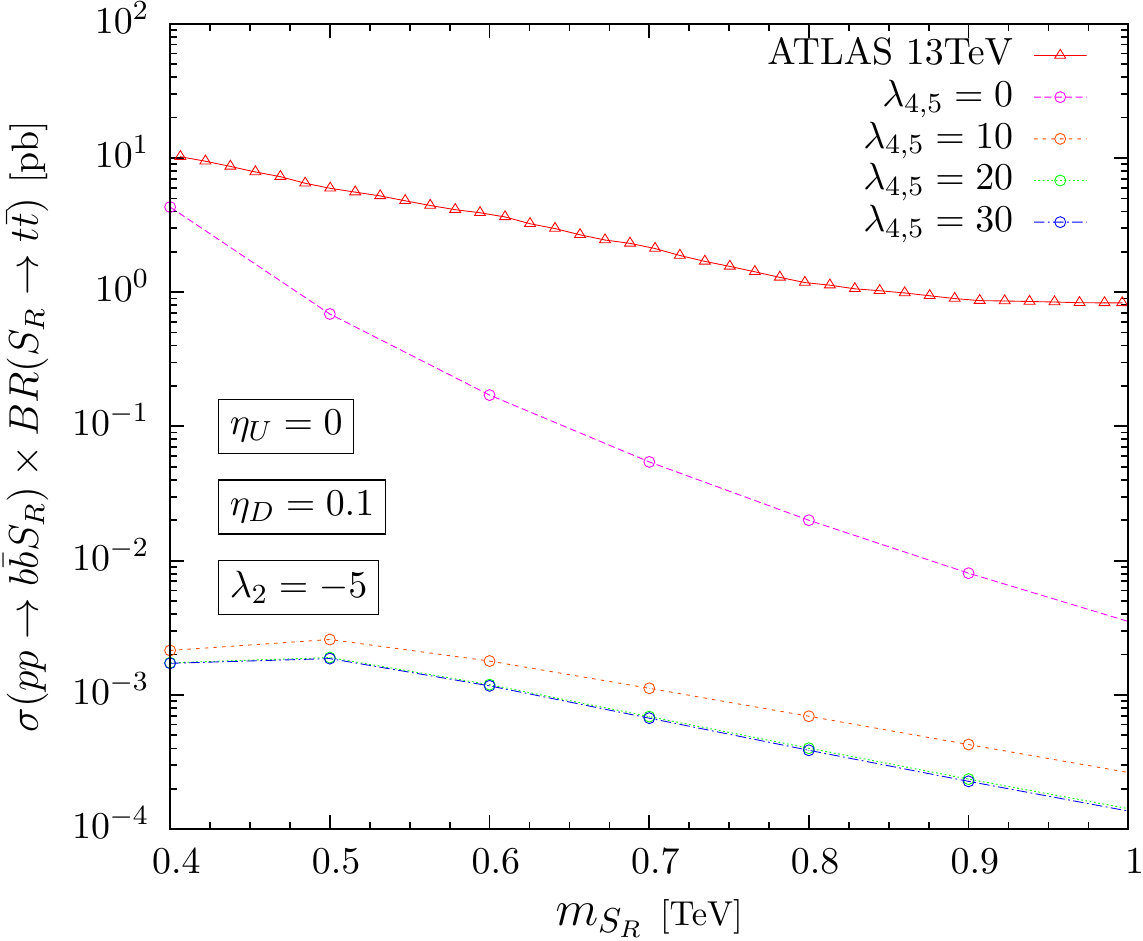} \hspace{0.5cm}
	\includegraphics[scale=0.57]{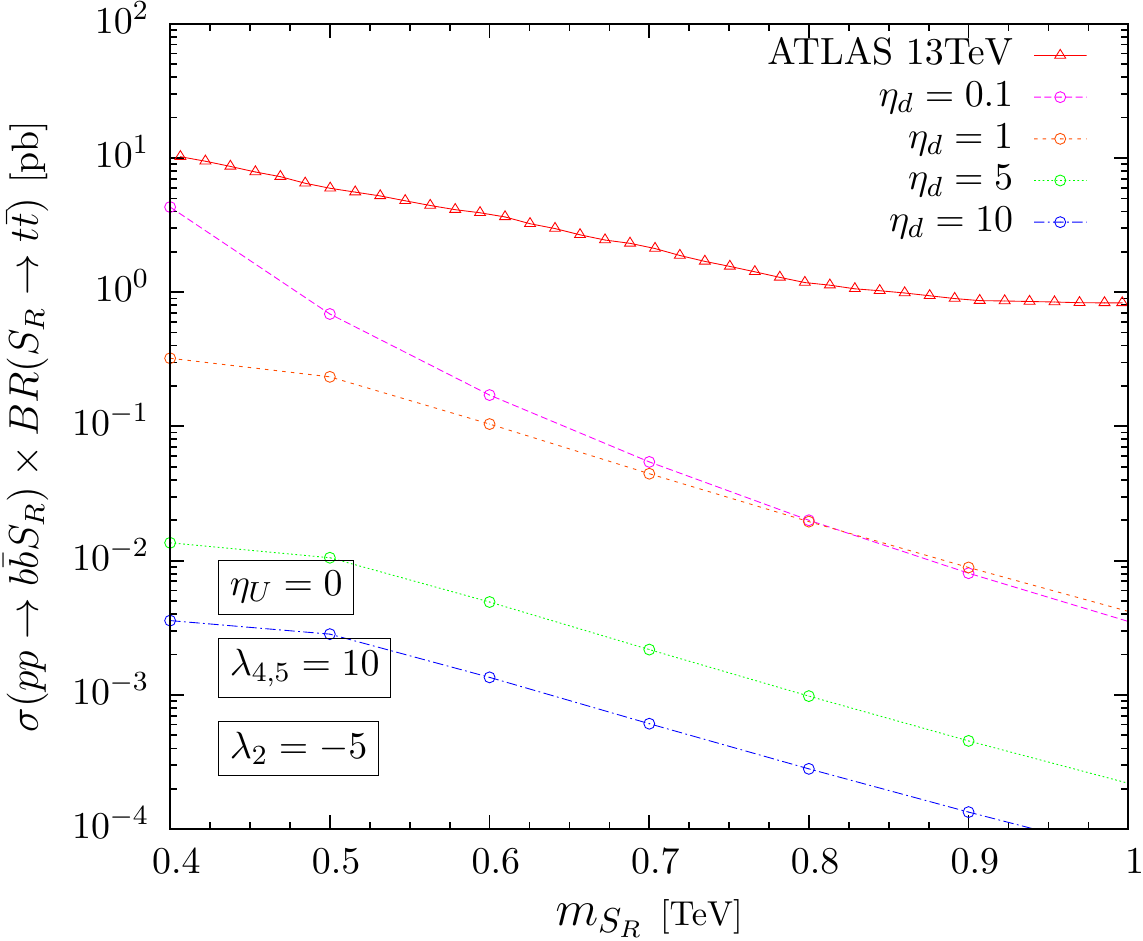}
	\caption{ Associated production of octet scalar ($S_R$) and its decay into 2 tops at 13TeV compared with Figure 21 of ATLAS preliminary results \cite{ATLAS:2016btu}(red curve). }
	\label{fig:2t2b}
\end{figure}

This channel is becoming accessible through searches for associated production of a new resonance with a pair of b-quarks from the preliminary results \cite{ATLAS:2016btu}. No constraints result from this case yet as illustrated in Figure~\ref{fig:2t2b}.

\section{Summary}

We have extracted constraints on the parameter space of the MW model by comparing the cross-sections for dijet, top-pair, dijet-pair, $t\bar t t \bar t$ and $b\bar b b \bar b$ production at the LHC with the strongest available experimental limits from ATLAS or CMS at 8 or 13 TeV. Our comparisons offer rough estimates as  the experimental collaborations have not considered this particular model and their results are somewhat model dependent, for example through the parton process assumed to produce the hypothetical resonance. 

Here we  summarize of our main findings within the two main assumptions: $\lambda_2 \sim -5$  to suppress decays involving leptons;  and $\lambda_4=\lambda_5 =10 $ to remain perturbative. 

\begin{figure}[h]
\centering
	\includegraphics[scale=0.5]{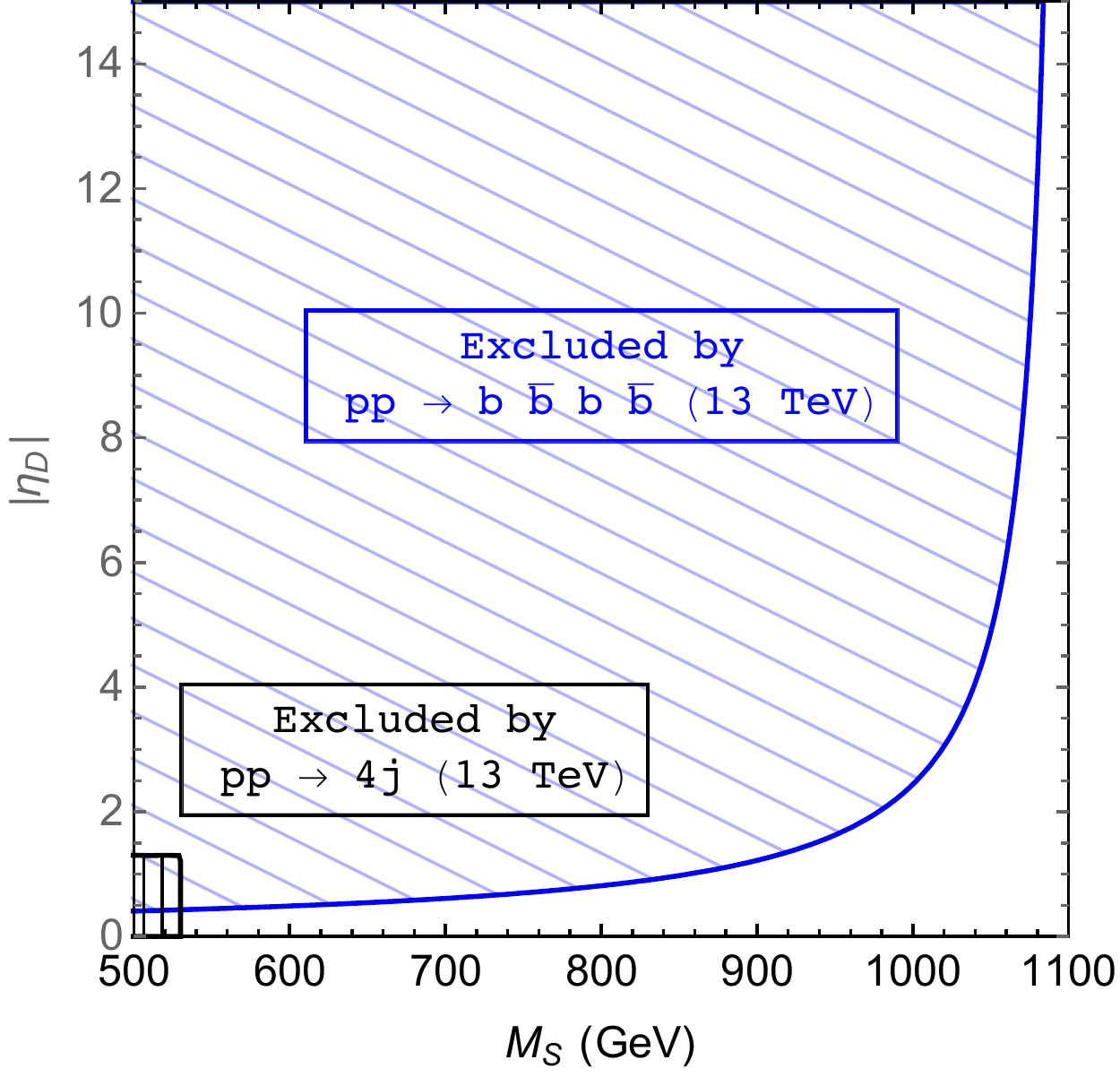} \hspace{0.5cm}
	\includegraphics[scale=0.5]{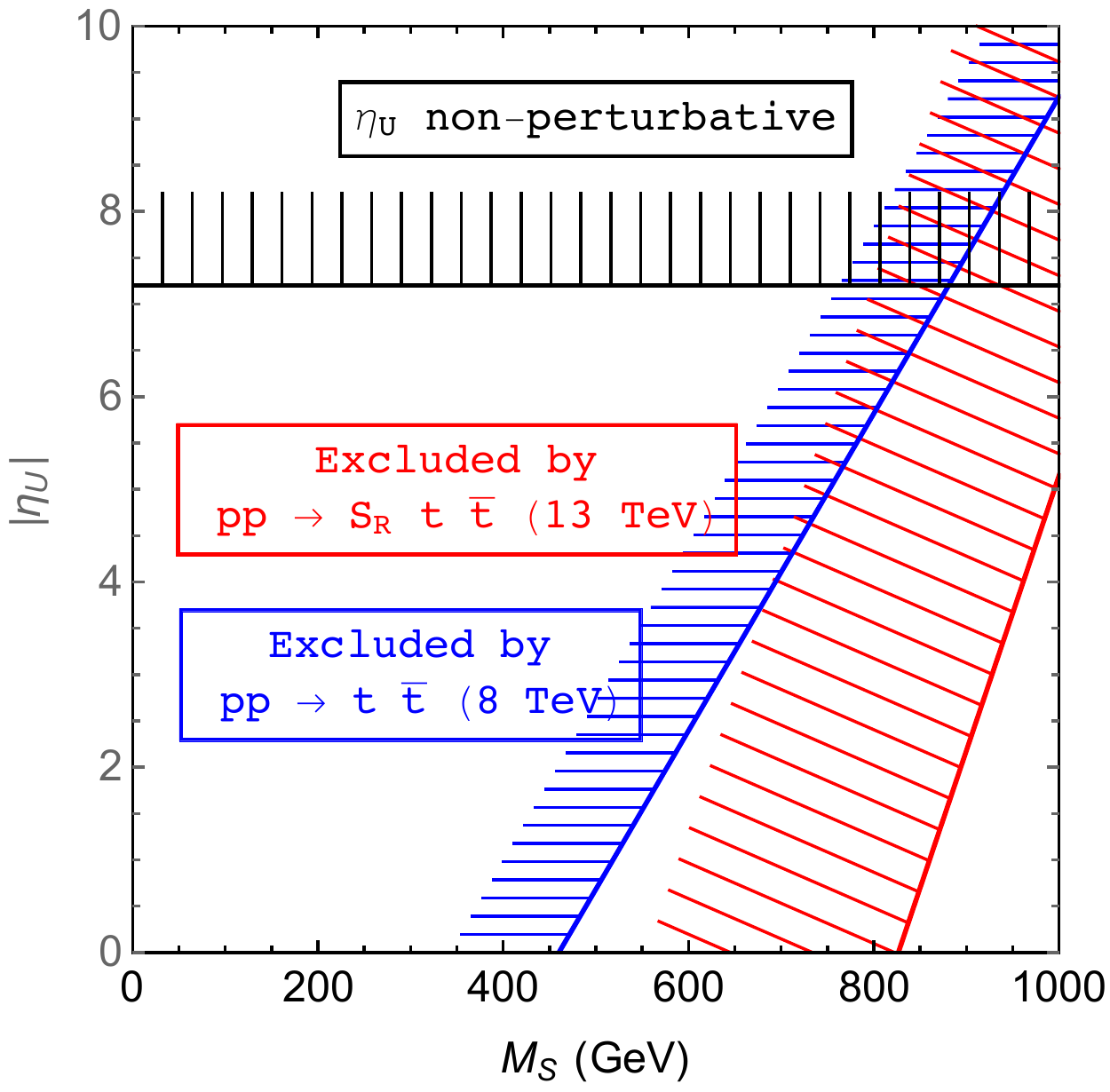}
	\caption{ Summary of constraints on $\eta_U,\eta_D,m_{S_R}$ parameter space for $\lambda_4=10$. The left panel corresponds to $\eta_U=0$ and the right panel to $\eta_D=0$.}
	\label{conetau}
\end{figure}

\begin{itemize}

\item Without couplings to up-type quarks, $\eta_U=0$, the best bound ranges from $m_S > 530 $~GeV to $m_S> 1060$~GeV for all values of $|\eta_D|$ as shown in Figure~\ref{conetau}. Figure~\ref{SvsSS} indicates that in the perturbative range for $\lambda_4$, $S_R$ pair production dominates over single $S_R$ production explaining why the most stringent bound  arises from production of two $b\bar{b}$ pairs.  The very low sensitivity to the value of $\eta_D$ when it gets above $\sim 6$ is due to the saturation of $B(S_R\to b\bar{b})\sim 1$ that occurs in this region of parameter space. For very low values of $\eta_D \lsim 1.3$, the dominant decay mode of $S_R$ is into two gluons and the constraint $m_S > 530 $~GeV is placed by four jet production.

\item Without couplings to down-type quarks, $\eta_D=0$, the best bound on $m_S$ is shown on the right panel of  shown in Figure~\ref{conetau}. It arises from the associated production of  $S_R$ with a top-quark pair at 13 TeV,  followed by $S_R$ decaying to a second top-quark pair. Constraints from $pp\to t\bar{t}$ at 8~TeV are almost competitive as shown in Figure~\ref{conetau}, suggesting they may become more important when this channel is studied at 13 TeV.

\end{itemize}

\end{document}